\documentclass[prb,twocolumn,showpacs,preprintnumbers]{revtex4}

\usepackage{graphicx}

\renewcommand{\Im}{{\rm Im}}

\newcommand{\laszlo}{}

\def \a{\alpha}
\def \o{\omega}    
\def \s{\sigma}    
\def \e{\epsilon}
   \def \d{\delta}

\def \L{\Lambda}
\def \O{\Omega}

\def \H{\hat{\cal H}}
\def \del{\partial}
\def \half{{1 \over 2}}
\def \dag{\dagger}

\def \ra{\rightarrow}
\def \up{\uparrow}
\def \down{\downarrow}

\def \be{ \begin{equation} }
\def \ee{ \end{equation} }
\def \bea{ \begin{eqnarray} }
\def \eea{ \end{eqnarray} }
\def \bea* { \begin{eqnarray*}}
\def \eea* { \end{eqnarray*} }

\def \l{\left}

\begin{document}
\title{Kondo quantum dot coupled to ferromagnetic leads: \\
 a study by numerical renormalization group technique}
\author{M.~Sindel,$^1$
  L.~Borda,$^{1,2}$
  J.~Martinek,$^{3,4,5}$
  R.~Bulla,$^{6}$
  J.~K\"onig,$^{7}$
  G.~Sch\"on,$^{^5}$
  S.~Maekawa,$^{3}$
  and J.~von~Delft$^1$}
\affiliation{
 $^1$Physics Department, Arnold Sommerfeld Center for
Theoretical Physics, and Center for NanoScience,
Ludwig-Maximilians-Universit\"at M\"unchen, 80333 M\"unchen,
Germany\\
 $^2$Research Group ``Theory of Condensed Matter'' of the Hungarian
Academy of Sciences, TU Budapest, H-1521, Hungary \\
 $^3$Institute for Materials Research, Tohoku University, Sendai 980-8577,
Japan \\
 $^4$Institute of Molecular
Physics, Polish Academy of Sciences, 60-179 Pozna\'n, Poland \\
 $^5$Institut f\"{u}r Theoretische
Festk\"{o}rperphysik and DFG-Center
    for Functional Nanostructures (CFN), Universit\"{a}t Karlsruhe,
    D-76128 Karlsruhe, Germany.\\
 $^6$Theoretische Physik III, Elektronische Korrelationen und
    Magnetismus, Universit\"at Augsburg, Augsburg, Germany\\
 $^7$Institut f\"ur Theoretische Physik III, Ruhr-Universit\"at
Bochum, 44780 Bochum, Germany
 }

\date{\today}

 \begin{abstract}
We systematically study the influence of ferromagnetic leads on
the Kondo resonance in a quantum dot tuned {\laszlo to} the local
moment regime. We employ Wilson's numerical renormalization group
method, extended to handle leads with a spin asymmetric density of
states, to identify the effects of (i) a finite spin polarization
in the leads (at the Fermi-surface), (ii) a Stoner splitting in
the bands (governed by the band edges) and (iii) an arbitrary
shape of the leads density of states. For a generic lead density
of states the quantum dot favors being occupied by a particular
spin-species due to exchange interaction with ferromagnetic leads
leading to a suppression and splitting of the Kondo resonance. The
application of a magnetic field can compensate this asymmetry
restoring the Kondo effect. We study both the gate-voltage
dependence (for a fixed band structure in the leads) and the spin
polarization dependence (for fixed gate voltage) of this
compensation field for various types of bands. Interestingly, we
find that the full recovery of the Kondo resonance of a quantum
dot in presence of leads with an {\laszlo energy dependent}
density of states is not only possible by an appropriately tuned
external magnetic field but also via an appropriately tuned gate
voltage. For flat bands simple formulas for the splitting of the
local level as a function of the spin polarization and gate
voltage are given.
\end{abstract}



\maketitle
\section{Introduction}
\label{Sec_Intro}

The interplay between different many-body phenomena, such as
superconductivity, ferromagnetism, or the Kondo effect, has
recently attracted a lot of experimental and theoretical
attention. A recent experiment of Buitelaar~\cite{schoenenberger}
{\it et al.} nicely demonstrated that Kondo correlations compete
with superconductivity in the leads. The interplay between Kondo
correlations and itinerant electron ferromagnetism in the
electrodes, has theoretically been intensively studied within the
last
years,~\cite{sergueev,zhang,bulka,lopez,martinek1,martinek2,choi,utsumi}
initially leading to controversial conclusions.

For effectively single-level quantum dots (i.e. dots with a level
spacing much bigger than the level broadening $\Gamma$), consensus
was found that a finite spin asymmetry in the density of states in
the leads results (in general) in a splitting and suppression of
the Kondo resonance. This is due the spin dependent broadening and
renormalization of the dot level position induced by
spin-dependent quantum charge fluctuations. In terms of the Kondo
spin model it can be treated as an effective exchange interaction
between a localized spin on the dot and ferromagnetic leads.
Moreover, it was shown that a strong coupling Kondo fixed point
with a reduced Kondo temperature can
develop~\cite{martinek1,martinek2} even though the dot is coupled
to ferromagnetic leads, given an external magnetic
field~\cite{martinek1} or electric field~\cite{martinek2}
(gate-voltage) is tuned appropriately. Obviously, in the limit of
fully spin polarized leads, when only one spin component is
present for energies close to Fermi surface (half-metallic leads),
the effective screening of the impurity {\it cannot} take place
any more and the Kondo resonance does {\it not} develop. A part of
these theoretical predictions have recently been confirmed in an
experiment by Pasupathy~\cite{Ralph}  {\it et al}. The presence of
ferromagnetic leads could also nicely explain the experimental
findings of Nyg{\aa}rd~\cite{nygaard} {\it et al}.

Since the interplay between ferromagnetism and strong correlation
effects is one of the important issues in spintronics
applications, there are currently many research activities going
on in this direction. The goal is to manipulate the magnetization
of a local quantum dot (i.e. its local spin) by means of an
external parameter, such as an external magnetic field or an
electric field (a gate voltage), with a high accuracy. This would
provide a possible method of writing information in a magnetic
memory.\cite{maekawa} Since it is extremely difficult to confine a
magnetic field such that it only affects the quantum dot under
study, it is of big importance to search for alternative
possibilities for such a manipulation (e.\ g.\ by means of a local
gate-voltage as proposed by the authors~\cite{martinek3}).

In this paper we push forward our previous work by performing a
systematic analysis on the dependence of physical quantities on
different band structure properties. Starting from the simplest
case we add the ingredients of a realistic model one-by-one
allowing a deeper understanding of the interplay of Kondo model
and itinerant electron ferromagnetism. In this paper we extend our
recent studies carried out in this
direction~\cite{martinek1,martinek2,martinek3} and illustrate the
strength of the analytical methods by comparing the results
predicted by them to the results obtained by the exact numerical
renormalization group (NRG) method.\cite{Wilson} While in
Refs.[\onlinecite{sergueev,zhang,bulka,lopez,martinek1,martinek2,choi,utsumi}]
the dot was attached to ferromagnetic leads with an unrealistic,
spin-independent and flat band - with a spin-dependent tunneling
amplitude - we generalize this treatment here by allowing for
arbitrary density of states (DOS) shapes.~\cite{martinek3} In
particular, we carefully analyze the consequences of typical DOS
shapes in the leads on the Kondo resonance. We explain the
difference between these shapes and provide simple formulas (based
on perturbative scaling analysis~\cite{Anderson,Haldane}) that
explain the numerical results analytically.

We study both the effect of a finite leads spin polarization {\it
and} the gate voltage dependence of a single-level quantum dot
contacted to ferromagnetic leads with three relevant DOS classes:
(i) for flat bands without Stoner-splitting, (ii) for flat bands
with Stoner-splitting and (iii) for an energy dependent DOS (also
including Stoner-splitting). For this sake we employ an extended
version of the NRG method to handle arbitrary shaped
bands.~\cite{Bulla:2004}

The article is organized as follows: In Sec.~\ref{Sec_Model} we
define the model Hamiltonian of the quantum dot coupled to
ferromagnetic leads. In Sec.~\ref{sec_NRG_mapping} we explain
details of the Wilson's mapping on the semi-infinite chain in the
case of the spin-dependent density of states with arbitrary energy
dependence. Using the perturbative scaling analysis we give
prediction for a spin-splitting energy for various band shapes in
Sec.~\ref{sec:scaling}. In Sec.~\ref{sec_flatband_noshift} the
results for spin-dependent flat DOS are demonstrated together with
the Friedel sum rule analysis. The effect of the Stoner splitting
is discussed in Sec.~\ref{sec_flatband_shift} together with
comparison to experimental results from Ref.[\onlinecite{nygaard}]
and an arbitrary band structure in Sec.~\ref{sec_arb_bands}. We
summarize our findings then in Sec.~\ref{sec_conclusion}.

\section{Model: quantum dot coupled to ferromagnetic leads}
\label{Sec_Model}

We model the problem at hand by means of a single-level dot of
energy $\epsilon_{\rm d}$ (tunable via an external gate-voltage
$V_{\rm G}$) and charging energy $U$ that is coupled to identical,
noninteracting leads (in equilibrium) with Fermi-energy $\mu=0$.
Accordingly the system is described by the following Anderson
{\laszlo impurity} model
\begin{eqnarray}
\H&=&\H_{\ell}+\H_{\ell d}+\H_{d} \; , \nonumber\\
\H_{d}&=&\epsilon_{\rm d} \sum_{\sigma} \hat n_\sigma + U
  \hat n_\uparrow \hat n_\downarrow  - B S_{z} \; ,
  \label{eq:Hd}
\end{eqnarray}
with the lead and the tunneling part of the Hamiltonian
\begin{eqnarray}
 \H_{\ell} &=& \sum_{rk \sigma} \epsilon_{rk \sigma}
c_{rk\sigma}^{\dagger}
  c_{rk \sigma} \; ,
  \label{eq:Hleads}
 \\
 \H_{\ell d} &=& \sum_{r k \sigma} (V_{rk} d_{\sigma}^{\dagger}
    c_{rk \sigma} + \textrm{h.c.}) \; .
      \label{eq:Htunneling}
\end{eqnarray}
Here $c_{rk\sigma}$ and $d_\sigma$ ($  \hat n_\sigma =
d_{\sigma}^{\dagger} d_{\sigma} $) are the Fermi operators for
electrons with momentum $k$ and spin $\sigma$ in lead $r$ ($r={\rm
L/R}$) and in the dot, respectively. The spin-dependent dispersion
in lead $r$, parametrized by  $\epsilon_{rk \sigma}$, reflects the
spin-dependent DOS,  $\rho_{r\s}(\omega)= \sum_{ k } \delta \l(
\omega - \epsilon_{rk \sigma} \right)$, in lead $r$; {\it all}
information about energy and spin dependency in lead $r$ is
contained in the dispersion function $\epsilon_{rk \sigma}$.
$V_{rk}$ labels the tunneling matrix-element between the impurity
and lead $r$, $S_{z} = (\hat n_\uparrow- \hat n_\downarrow)/2$,
and the last term in Eq.~(\ref{eq:Hd}) denotes the Zeeman energy
due to external magnetic field $B$ acting on the dot spin only.
Here we neglect the effect of an external magnetic field on the
leads' electronic structure as well as a stray magnetic field from
the ferromagnetic leads. The coupling between the dot level and
electrons in lead $r$ leads to a broadening and a shift of the
level $\epsilon_{\rm d}$, $\epsilon_{\rm d}\rightarrow
\tilde{\epsilon}_{\rm d} $ (where the tilde denotes the
renormalized level). {\laszlo The {\it energy} and {\it spin}
dependency of the broadening and the shift,
determined by the coupling, 
$\Gamma_{r\s}(\o)=\pi\rho_{r\s}(\o)|V_{r}(\o)|^2$,
plays the key role in the effects outlined in this paper.}
Henceforth, we assume $V_{rk}$ to be real and $k$-independent,
$V_{rk\s}=V_r$, and lump all energy and spin-dependence of
$\Gamma_{r\s}(\o)$ into the DOS in lead $r$,
$\rho_{r\s}(\o)$.~\cite{Normalization}

%
%
%
%
Without loosing generality we assume the coupling to be symmetric,
$V_{\rm L} = V_{\rm R}$. 
Accordingly,
by performing a unitary transformation~\cite{Glazman:1988}
 $\H_{\ell}$ simplifies to
$\H_{\ell}=\half\sum_{k\sigma}\l(\epsilon_{{\rm L} k
\sigma}+\epsilon_{{\rm R} k \sigma}\right)
\a_{sk\s}^{\dag}\a_{sk\s}$,
where $\a_{sk\s}$ denotes the proper unitary combination of lead operators
which couple to the quantum dot and we dropped the part of the 
lead Hamiltonian which is decoupled from the dot.
With the help of the definitions $V\equiv\sqrt{V_{L}^2+V_{R}^2}$,
$\a_{k \sigma}\equiv \a_{sk \sigma}$ and $\e^*_{k \sigma}
\equiv\half\l(\epsilon_{Lk \sigma}+\epsilon_{Rk \sigma}\right)$
the full Hamiltonian can be cast into a compact form
\begin{eqnarray}
\H = \sum_{k \sigma} \e^*_{k \sigma}\a_{k\sigma}^{\dagger}
  \a_{k \sigma} + \sum_{ k \sigma} V\l( d_{\sigma}^{\dagger}
    \a_{k \sigma} + \textrm{h.c.}\right) +\H_d \:  ,
    \label{eq:AMf}
\end{eqnarray}
with $\H_d$ as given in Eq.~(\ref{eq:Hd}).

\subsection{Ferromagnetic leads}

 For ferromagnetic materials electron-electron interaction in the
leads give rise to magnetic
 order and spin-dependent DOS,
 $\rho_{r\uparrow}(\omega) \neq \rho_{r\downarrow}(\omega)$.
Magnetic order of typical band ferromagnets like Fe, Co, and Ni is
mainly related to electron correlation effects in the relatively
narrow $3d$ sub-bands, which only weakly hybridize with $4s$ and
$4p$ bands \cite{nolting}. We can assume that due to a strong
spatial confinement of $d$ electron orbitals, the contribution of
electrons from $d$ sub-bands to transport across the tunnel
barrier can be neglected \cite{tsymbal}. In such a situation the
system can be modeled by noninteracting \cite{interaction} $s$
electrons, which are spin polarized due to the exchange
interaction with uncompensated magnetic moments of the completely
localized $d$ electrons. In mean-field approximation one can model
this exchange interaction as an effective molecular field, which
removes spin degeneracy in the system of noninteracting conducting
electrons, leading to a spin-dependent DOS.

{\it Parallel and antiparallel leads' magnetization. --}
In experiments very frequently the electronic transport
measurements are performed for two configurations of the leads'
magnetization direction~\cite{Ralph}: the parallel and
antiparallel alignment. By comparison of electric current for
these two configurations one can calculate the tunneling
magnetoresistance (TMR) important parameter for the application of
the magnetic tunnel junction~\cite{maekawa}.

 In this paper we restrict the leads' magnetization direction to be either
(i) {\it parallel}, i.e. the left and right lead have the DOS $
\rho_{\rm L \s}(\o) $ and $ \rho_{\rm R \s}(\o) $ respectively, so
the total DOS corresponding to the total dispersion $\e^*_{k \s}$
($\forall k \in \l[-D_0;D_0\right]$) is given by  $\rho_{\s}(\o) =
\rho_{\rm L \s}(\o)+\rho_{\rm R \s}(\o) $.
(ii) {\it antiparallel}, the magnetization direction of one of the
leads (let us consider the right one) is reverted so $ \rho_{\rm R
\s}(\o) \rightarrow \rho_{\rm R \bar{\s}}(\o)$, where
$\bar{\s}=\down(\up)$ if $\s=\up(\down)$, so the total DOS is
described then by $\rho_{\s}(\o) = \rho_{\rm L \s}(\o)+\rho_{\rm R
\bar{\s}}(\o) $. Here $D_0$ labels the full (generalized)
bandwidth of the conduction band (further details can be found
below). Effects related to leads with non-collinear leads'
magnetization are not discussed here~\cite{Koenig}.

For the special case of both leads made of the same material in
the parallel alignment $ \rho_{\rm L \s}(\o) = \rho_{\rm R
\s}(\o)$ and in the antiparallel alignment $ \rho_{\rm L \s}(\o) =
\rho_{{\rm R} \bar{\s}}(\o)$. Therefore for the antiparallel case
it gives the total DOS  to be spin independent $ \rho_{ \up}(\o) =
\rho_{\rm L \up}(\o)+ \rho_{\rm R \down}(\o)= \rho_{\rm L
\down}(\o)+ \rho_{\rm R \up}(\o) = \rho_{ \down}(\o)$. In a such
situation one can expect the usual Kondo effect as for normal
(non-ferromagnetic) metallic leads, however, the conductance will
be diminished due to mismatch of the density of states described
by the prefactor of the integral in Eq.~(\ref{spin_res_Cond})

\subsection{Different band structures}

In this paper we will consider different type of total
spin-dependent band structures $\rho_{\s}(\o)$ independently of
the particular magnetization direction of leads. The particular
magnetization configuration will affect only the linear
conductance $G_{\s}$ due to the DOS mismatch for both leads as
discussed in detail later.

\subsubsection{Flat band}

The simplest situation, where one can account spin asymmetry in is
a flat band with the energy-independent DOS
$\rho_{\s}(\o)=\rho_{\s}$.
Then the spin asymmetry can be parameterized just by single
parameter the spin imbalance in the DOS at the Fermi energy $\o
=0$. For flat bands the knowledge of the spin-polarization $P$ of
the leads, defined as
 \be
 \label{Def_P}
 P=\frac{\rho_{\up}(0)-\rho_{\down}(0)}
{\rho_{\up}(0)+\rho_{\down}(0)}\; ,
 \ee{equation}
is sufficient to fully parameterize $\rho_{\s}(\o)$ (see
Fig.~\ref{fig:DOS_flat_noshift}). This particular DOS shape is a
special due to the fact, that the particle-hole symmetry is
conserved in the electrodes leading to particular behaviour for
the symmetric Anderson model. This type model of the leads will be
considered in Sec.~\ref{sec_flatband_noshift}.

\subsubsection{Stoner splitting}

One can generalize this model and break the particle-hole symmetry by
taking
the Stoner splitting
into account. A consequence of the conduction
electron ferromagnetism is that the spin-dependent bands are
shifted relative to each other (see Fig.~\ref{fig:DOS_arbitrary}).
For a finite value of that shift $\Delta_{\s}$, the spin-$\s$ band
ranges between $-D+\Delta_{\s}\leq \omega \leq D+\Delta_{\s}$
(with the 'original' bandwidth $D$). In the limit $\Delta_{\s}=0$
only energies within the interval $[-D;D]$ are considered, a
scheme usually used in the NRG calculations~\cite{martinek2}. This
relative shift between the two spin-dependent bands leads to so
called Stoner splitting $\Delta$, defined as
\be
 \label{Stoner_splitting}
 \Delta=\Delta_{\down}-\Delta_{\up}
\ee
at the band edges of the conduction band (see
Fig.~\ref{fig:DOS_Flat_constshift}). The flat band model with a
Stoner splitting and consequences of the particle-hole symmetry
braking is considered in Sec.~\ref{sec_flatband_shift}.

\subsubsection{Arbitrary band structure}

\begin{figure}[t!]
\centering
\includegraphics[width=1.0\columnwidth]{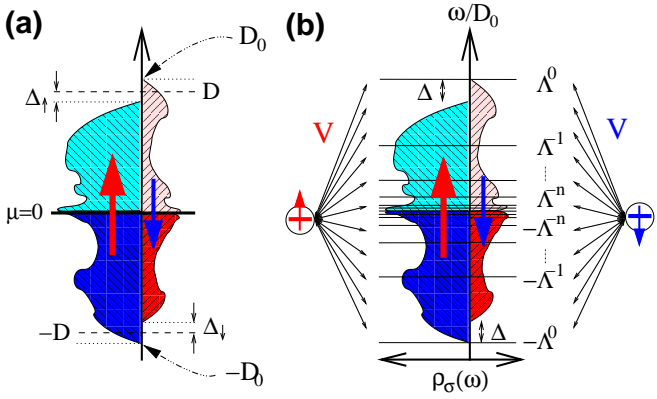}
\caption{ (a) Example of an {\it energy}- and {\it spin}-dependent
leads DOS $\rho_{\s}(\omega)$ with an additional spin-dependent
shift $\Delta_{\s}$. To perform the logarithmic discretization a
generalized bandwidth $D_0$ is defined. Since we allow for bands
with energy and spin dependence the discretization is performed
for each spin-component separately, see panel (b). Since $\H_{\ell
d}$ does not include spin-flip processes an impurity electron of
spin $\s$ [circles in panel (b)] couples to lead electrons of spin
$\s$, $\s=\up (\down)$, only. Panel (b) also illustrates that
impurity electrons couple to leads' electron and hole sates with
{\it arbitrary} energy $\omega$, $|\omega|\leq D_0$.}
 \label{fig:DOS_arbitrary}
\end{figure}

Since the spin-splitting of the dot level is determined by the
coupling to all occupied and unoccupied (hole) electronic states
in the leads, the shape of the whole band plays an important role.
Therefore it is reasonable to consider arbitrary DOS shape, which
cannot be parameterized by a particular set of parameters as the
imbalance between $\up$- and $\down$-electrons at the Fermi energy
or a Stoner splitting (see Fig.~\ref{fig:DOS_arbitrary}).
Therefore in Sec.~\ref{sec_arb_bands} we will consider a model
with a more complex band structure and in the next
Sec.~\ref{sec_NRG_mapping} we will develop the NRG technique for
arbitrary band structure.

\section{The method: NRG for arbitrary spin-dependent density of states}
\label{sec_NRG_mapping}

In our analysis we take the ferromagnetic nature of the
noninteracting leads by means of a {\it spin-} and {\it
energy-dependent} DOS $\rho_{\s}(\omega)$ into account. A general
example is given in Fig.~\ref{fig:DOS_arbitrary}.
To compute the properties of the model described above we have
extended numerical renormalization group (NRG) technique
calculation to handle spin-dependent density of states. In order
to understand to what extent the method applied here is different
from the standard NRG it is adequate to briefly review the general
concepts of NRG.

The NRG technique was invented by Wilson in the 70's to solve the
Kondo problem\cite{Wilson} -- later it was extended to handle
other quantum impurity models as
well\cite{Krishna-murthy,Costi,Hofstetter,KM_mapping}. In his
original work Wilson considered a spin-independent flat density of
states for the conduction electrons. Closely following
Refs[~\onlinecite{Bulla:1997,Bulla:2004}] --where the mapping for
the case of energy-dependent DOS was given-- we generalize that
procedure for the case of the Hamiltonian given in
Eq.~(\ref{eq:AMf}) which contains leads with an {\it energy}- and
{\it spin}-dependent DOS.

It is convenient to bring the Hamiltonian given by
Eq.~(\ref{eq:AMf}), into a continuous
representation~\cite{Bulla:1997} before the generalized mapping is
started. The replacement of the discrete fermionic operators by
continuous ones, $\a_{k\s}\rightarrow\a_{\o\s}$, translates the
lead and the tunneling part of the Hamiltonian into
\begin{eqnarray}
\H_{\ell} &=& \sum_{\s}\int_{-1}^{1}d\o \;g_{\s}(\o)\;
\a^{\dag}_{\o\s}\a_{\o\s}
 \\
 \H_{\ell d} &=& \sum_{\s}\int_{-1}^{1}d\o\;h_{\s}(\o)\;
\l(d^{\dag}_{\s}\a_{\o\s} + \textrm{h.c.}\right) \; .
    \label{eq:Hcontinuous}
\end{eqnarray}
As shown in Ref.~\onlinecite{Bulla:1997} the hereby defined
generalized {\it dispersion} $g_{\s}(\o)$ and {\it hybridization}
$h_{\s}(\o)$ functions have to satisfy the relation \be
\label{equal_action} \frac{\del g^{-1}_{\s}(\o)} {\del
\o}\l[h_{\s}\l(g^{-1}_{\s}(\o)\right)\right]^2=
\rho_{\s}(\o)\l[V_{\s}(\o)\right]^2, \ee where $g^{-1}_{\s}(\o)$
is the inverse of $g_{\s}(\o)$, what ensures that the action on
the impurity site is identical both in the discrete and the
continuous representation. Note that there are many possibilities
to satisfy Eq.~(\ref{equal_action}).
%
%

The key idea of Wilson's NRG ~\cite{Wilson} is a logarithmic
discretization of the conduction band, by introducing a
discretization parameter $\Lambda$, which defines energy intervals
$]-D_0\Lambda^{-n};-D_0\Lambda^{-n-1}]$ and
$[D_0\Lambda^{-n-1};D_0\Lambda^{-n}[$ in the conduction band
($n\in \mathbf{N}_0$). Within the $n$-th interval of width
 $d_n=\L^{-n}\l(1-\L^{-1}\right)$ a Fourier expansion of
the lead operators $\Psi_{np}^{\pm}(\o)$
with fundamental frequency  $\O_n= 2\pi/d_n$ is defined
\begin{eqnarray*}
\Psi_{np}^{\pm}(\o)=
\cases{\frac{1}{\sqrt{d_n}}e^{\pm i\O_n p\o} &\textrm{if}
$\Lambda^{-(n+1)}\leq \pm \o <\Lambda^{-n}$ \cr
0 &\textrm{else} \cr }
\end{eqnarray*}
Here the subscripts $n$ and $p$ ($\in\mathbf{Z}$) label the
corresponding interval and the harmonic index, respectively, while
the superscript marks positive ($+$) or negative ($-$) intervals,
respectively.

The above defined Fourier series now allows one to replace the
continuous fermionic conduction band operators $a_{\o\s}$
by discrete ones $a_{np\s}$ ($b_{np\s}$) of harmonic index $p$ and spin $\s$ acting on the
$n$-th positive (negative) interval only
\be \label{Fourier_CB}
\a_{\o\s}=\l\{\sum_{np}\l[a_{np\s}\Psi_{np}^{+}(\o)+
b_{np\s}\Psi_{np}^{-}(\o)\right]\right\}. \ee
Impurity electrons couple {\it only}  to the $p=0$ mode of the
lead operators, given the energy-dependent generalized
hybridization $h_{\s}(\o)$ is replaced by a constant
hybridization, $h_{\s}(\o)\ra h_{n\s}^{+}$ for $\o>0$ (or
$h_{n\s}^{-}$ for $\o<0$, respectively). Obviously, the particular
choice of constant hybridization $h_{n\s}^{\pm}$ demands the
generalized dispersion $g_{\s}(\o)$ to be adjusted accordingly,
such that Eq.~(\ref{equal_action}) remains valid. Details of this
procedure can be found in Appendix~\ref{aver_hybr}. Since we adopt
this strategy the harmonic index $p$ (the impurity couples only to
lead operators of harmonic index $p=0$) will be
{\laszlo dropped below}.

Defining a fermionic operator \be\label{f_max_localized}
f_{0\s}\equiv
\frac{1}{\sqrt{\xi_{0\s}}}\sum_n\l(a_{n\s}\gamma_{n\s}^+\;+\;
b_{n\s}\gamma_{n\s}^-\right), \ee with $\xi_{0\s}=\sum_n
\l[\l(\gamma_{n\s}^+\right)^2+
\l(\gamma_{n\s}^-\right)^2\right]=\int_{-1}^{1}\Gamma_{\s}(\o)d\o$
and the coefficients $\gamma_{n\s}^{\pm}$ as given in
Appendix~\ref{aver_hybr}~\cite{remark_1}, reveals that the
impurity effectively couples to a {\it single} fermionic degree of
freedom only, the zeroth site of the Wilson chain [for further
details see Eq.~(\ref{App_tunn_comp})].
 Therefore the tunneling
part of $\H$ can be written in a compact form as \be
\label{final_tunneling_NRG} \H_{\ell
d}=\sum_{\s}\l[\sqrt{\frac{\xi_{0\s}}{\pi}} \l(
d^{\dag}_{\s}f_{0\s}\;+\;\textrm{h.c.} \right)\right]. \ee The
final step in the NRG-procedure is the transformation of the
conduction band $\H_{\ell}$ into the form of a linear chain. This
goal is achieved via the tridiagonalization procedure developed by
L\'anczos~\cite{Lanczos:1950} \be \label{lead_Wilson_chain}
\H_{\ell}=  \sum_{\sigma n=0}^\infty [
   \varepsilon_{n\sigma} f_{n\sigma}^\dagger f_{n\sigma}
+ t_{n\sigma} ( f_{n\sigma}^\dagger f_{n+1\sigma}
 + f_{n+1\sigma}^\dagger f_{n\sigma})].
\ee
In general the on-site energies $\varepsilon_{n\s}$ and hopping matrix elements
$t_{n\s}$ along the Wilson-chain need to be determined numerically.
Besides the matrix elements $\e_{n\s}$ and $t_{n\s}$ coefficients
$u_{nm\s}$ and $v_{nm\s}$, which define the
fermionic operators $f_{n\sigma}$
\be \label{Ansatz_f_NRG}
f_{n\s}\equiv\sum_{m=0}^{\infty}\l(u_{nm\s}a_{m\s}+v_{nm\s}b_{m\s}
\right)
\ee
already used in Eq.~(\ref{lead_Wilson_chain}), need to be determined.
One immediately anticipates from Eq.~(\ref{f_max_localized})
\be
u_{0m\s}=\gamma_{m\s}^+/\sqrt{\xi_{0\s}} \; , \;\;\;
v_{0m\s}=\gamma_{m\s}^-/\sqrt{\xi_{0\s}} \; .
 \label{eq:coef}
\ee
Equations which determine the matrix elements $\e_{n\s}$ and
$t_{n\s}$ and the coefficients $u_{nm\s}$ and $v_{nm\s}$ are given
in Appendix~\ref{app_CB}. Note that the on-site energies
$\varepsilon_{n\s}$ vanish in the presence of particle-hole
symmetry in the leads.

To summarize: Hamiltonians as the one given in Eq.~(\ref{eq:AMf})
can be cast into the form of a linear chain $\H_{\rm{LC}}$
\begin{eqnarray}
  &\H_{{\rm{LC}}}& =   \H_d
   + \sqrt{ {\xi_{0\sigma}}/{\pi}}\sum_{\sigma} [
  d^\dagger_{\sigma}f_{0\sigma} +
   f^\dagger_{0\sigma}d_{\sigma}  ]   \label{eq:H_si} \\
   & +& \sum_{\sigma n=0}^\infty [
   \varepsilon_{n\sigma} f_{n\sigma}^\dagger f_{n\sigma}
+ t_{n\sigma} ( f_{n\sigma}^\dagger f_{n+1\sigma}
 + f_{n+1\sigma}^\dagger f_{n\sigma})] \; ,
  \nonumber
\end{eqnarray}
 even though one is dealing with energy and spin-dependent leads.
In general, however, this involves
numerical determination of the matrix elements
$\varepsilon_{n\sigma}$ and $t_{n\sigma}\;$~\cite{Bulla:2004} in
contrast to Krishna-murthy~\cite{Krishna-murthy} {\it et al.} (who
considered flat bands) no closed analytical expression for those
matrix elements is known.

Eq.~(\ref{eq:H_si}) nicely illustrates the strength of the NRG-procedure:
 As a consequence of
the energy separation guaranteed by the logarithmic discretization,
the hopping rate along the chain decreases as $t_n\sim\Lambda^{-n/2}$
(the on-site energies decays even faster)
which allows
us to diagonalize the chain Hamiltonian iteratively and in every
iteration to keep the states with the lowest lying energy eigenvalues
as the most relevant ones. This very fact underlines that
this method does not rely on any assumptions concerning leading
order divergences.~\cite{Hofstetter:2002}

\section{Perturbative scaling analysis}
\label{sec:scaling}

We can understand the spin-splitting of the spectral function
using Haldane's scaling approach\cite{Haldane}, where quantum
charge fluctuations are integrated out. The behavior discussed in
this paper can be explained as an effect of spin-dependent quantum
charge fluctuations, which lead to a spin-dependent
renormalization of the dot's level position $
\tilde{\epsilon}_{\rm d \sigma}$ and a spin-dependent level
broadening $\Gamma_{\sigma}$, which in turn induce spin splitting
of the dot level and the Kondo resonance.
Within this approach a spin-splitting of the local dot level, $
\Delta \epsilon_{\rm d} \equiv \delta\epsilon_{\rm d \uparrow} -
\delta\epsilon_{\rm d \downarrow} + B $, which depends on the {\it
full} band structure of the leads is obtained~\cite{martinek3}
where
\begin{eqnarray}
\delta\epsilon_{\rm d \sigma} \simeq -\frac{1}{\pi} \int d\omega
\left\{ \frac{\Gamma_{\sigma}(\omega)[1-f(\omega)]}{\omega -
\epsilon_{\rm d \sigma}} + \frac{\Gamma_{\bar {
\sigma}}(\omega)f(\omega)}{\epsilon_{\rm d \bar{
\sigma}}+U-\omega}
 \right\} \, .
  \label{splitt_gen}
\end{eqnarray}
Note that the splitting is {\it not} only determined by the leads
spin polarization $P$, i.\ e.\ the splitting is not only a
property of the Fermi surface. Eq.~(\ref{splitt_gen}) is the key
equation to explain the physics of the (spin-dependent) splitting
of the local level $\e_{\rm d \sigma}$. This equation explains the
spin-dependent occupation and consequently the splitting of the
spectral function of a dot that is contacted to leads with a
particular band-structure.
The first term in the curly brackets corresponds to electron-like
processes, namely charge fluctuations between a single occupied
state $ |\sigma \rangle $ and the empty $ |0 \rangle $ one, and
the second term to hole-like processes, namely charge fluctuations
between the states $ |\sigma \rangle $ and $ |2 \rangle $.
The amplitude of the charge fluctuations is proportional to $
\Gamma $, which for $ \Gamma \! \gg \!\! T $ determines the width
of dot levels observed in transport.
%

The exchange field given by Eq.~(\ref{splitt_gen}) gives rise to
precession of an accumulated spin on the quantum dot attached to
leads with non-collinear leads' magnetization are not
discussed here~\cite{Koenig}.

\subsection{Flat band}

Eq.~(\ref{splitt_gen}) predicts that even for systems with
spin-asymmetric bands $ \rho_{\uparrow}(\omega) \neq
\rho_{\downarrow}(\omega) $, the integral can give $ \Delta
\epsilon = 0 $, which corresponds to a situation where the
renormalization of $ \epsilon_{\rm d \sigma} $ due to
electron-like processes are compensated by hole-like processes. An
example is a system consisting of particle-hole symmetric bands, $
\rho_{ \sigma }(\omega) = \rho_{ \sigma }(- \omega )$, where no
splitting of the Kondo resonance ($ \Delta \epsilon_{\rm d} = 0 $)
for the symmetric point, $ \epsilon_{\rm d} = - U/2 $, appears.

For a flat band $ \rho_{\sigma}( \omega ) = \rho_{\sigma} $,
Eq.~(\ref{splitt_gen}) can be integrated analytically. For $ D_0
\gg U$, $ |\epsilon_{\rm d} | $ one finds:
$ \Delta \epsilon \simeq \left(  P \; \Gamma / \pi\right)
\mathrm{Re} [ \phi(\epsilon_{\rm d}) - \phi(U+ \epsilon_{\rm d})
 ] $,
where $ \phi(x) \equiv \Psi ( \frac{1}{2} + i {x}/{2 \pi T} ) $
and $\Psi(x)$ denotes the digamma function. For $ T = 0 $, the
spin-splitting is given by
\begin{eqnarray}
\Delta \epsilon_{\rm d} \simeq  \frac{P \; \Gamma }{\pi} \ln
\left( \frac{| \epsilon_{\rm d} |}{| U + \epsilon_{\rm d} |}
\right)  \; ,
  \label{Bcomp_dep}
\end{eqnarray}
showing a logarithmic divergence for $ \epsilon_{\rm d}
\rightarrow 0 $ or $ U + \epsilon_{\rm d} \rightarrow 0$.

\subsection{Stoner splitting}

\begin{figure}[t!]
\centering
\includegraphics[width=.350\columnwidth]{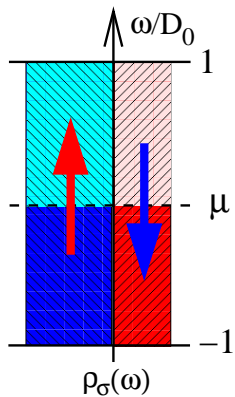}
\caption{The leads density of states of a flat [$
\rho_{s}(\omega)= \rho_{\s}$] but spin-dependent ($\rho_{\up} \neq
\rho_{\down}$) band for spin polarization $P=0.2$. The dark region
marks the filled states below the Fermi energy. Since
$\Delta_{\up}=\Delta_{\down}=0$ the generalized bandwidth
$D_0=D$.}
 \label{fig:DOS_flat_noshift}
\end{figure}

For real systems p-h symmetric bands cannot be assumed, however
the compensation  $ \Delta \epsilon = 0 $ using a proper tuning of
the gate voltage $\epsilon_{\rm d}$ is still possible.
We can analyze also the effect of the Stoner splitting by
considering a flat-band structure in
Fig.~\ref{fig:DOS_flat_noshift} using the value of the Stoner
splitting $\Delta = 0.2 (\frac{D}{D_0}) D_0$. Then also from
Eq.~(\ref{splitt_gen}) we can expect an additional spin-splitting
of the dot level induced by the presence of the Stoner field in
the leads even for spin polarization $P=0$ given by
\begin{eqnarray}
\Delta \epsilon_{\rm d}^{(\rm St)} \simeq  \frac{ \Gamma }{2 \pi}
\ln \left[  \frac{ (-\epsilon_{\rm d} +D_0-\Delta)(U +
\epsilon_{\rm d} +D_0 -\Delta)}{ (-\epsilon_{\rm d} +D_0)(U +
\epsilon_{\rm d} +D_0)} \right] \; ,
  \label{Bcomp_Stoner}
\end{eqnarray}
which for the symmetric level position, $\epsilon_{\rm d}=-U/2$,
it leads to
\begin{eqnarray}
\Delta \epsilon_{\rm d}^{(\rm St)} \simeq  \frac{ \Gamma }{ \pi}
\ln \left[  \frac{ U/2 + \epsilon_{\rm d} +D_0 -\Delta}{ U/2 +
\epsilon_{\rm d} +D_0} \right] \; ,
  \label{Bcomp_Stoner_symm}
\end{eqnarray}

Eq.~(\ref{Bcomp_Stoner}) also shows that the characteristic energy
scale of the spin-splitting is given by $ \Gamma $ rather than by
the Stoner splitting $ \Delta $ ($ \Delta \gg \Gamma $), since the
states far from the Fermi surface enter Eq.~(\ref{splitt_gen})
only with a logarithmic weight. The difference can be as large as
three orders of the magnitude, so in metallic ferromagnet the
Stoner splitting energy is of the order $\Delta \sim 1 \; eV$ but
still the effective molecular filed $ \Delta \epsilon_{\rm
d}^{(\rm St)} $ generated by it is a small fraction of $\Gamma$ -
of order of $ 1 \; meV $ so comparable with the Kondo energy scale
for molecular single-electron transistors.~\cite{Ralph} However,
the Stoner splitting introduces a strong p-h asymmetry, so it can
influence the character of gate voltage dependence significantly.

In the next Section we analyze the effect of different type of the
band structure using numerical renormalization group technique and
compare it to that obtained in this Section by scaling procedure.

\section{Flat bands without Stoner splitting}
\label{sec_flatband_noshift}

We start our analysis by considering normalized flat bands without
the Stoner splitting (i.\ e.\ $D_0=D$), as sketched in
Fig.~\ref{fig:DOS_flat_noshift}, with finite spin-polarization $P
\neq 0 $ [defined via Eq.~(\ref{Def_P})].\cite{DOS_flat_band} In
this particular case -- of spin- (but not energy-) dependent
coupling $\Gamma_{\sigma} \equiv \pi \rho_{\sigma} V^2$ -- the
coupling $\Gamma_{\sigma}$ can be parametrized via $P$,
$\Gamma_{\uparrow ( \downarrow ) } = {1 \over 2} \Gamma (1 \pm P)$
[here $+$ ($-$) corresponds to spin $\uparrow$ ($\downarrow$)],
 where $\Gamma = \Gamma_\uparrow +
\Gamma_\downarrow$.
Leads with a DOS as the one analyzed in this section have for
instance been studied in Ref.[\onlinecite{martinek2,choi}].

Since $\Gamma_{\s} \equiv \pi \rho_{\s}V^2$ the spin-dependence of
$\Gamma_{\s}$ can be absorbed by replacing $V\ra
V_{\s}=V\sqrt{\half(1\pm P)}$ in Eq.~(\ref{eq:AMf}), i.\ e.\ a
spin-dependent {\it hopping} matrix element,
 while treating the leads as unpolarized ones ($\rho_{\s}\ra \rho$).
This procedure has the particular advantage that the 'standard'
NRG procedure~\cite{Krishna-murthy} can be applied meaning that
the on-site energies (tunneling matrix elements), defined in
Eq.~(\ref{lead_Wilson_chain}), along the Wilson chain
$\varepsilon_{n\s}$ vanish, while $t_{n\sigma}$'s  turn out to be
{\it spin-independent}. Therefore it does not involve the solution
of the tedious equations given in Appendix~\ref{app_CB}.

\subsection{Spin and charge state}
 \label{sec:occupation}

\begin{figure}[t!]
\centering
\includegraphics[width=0.80\columnwidth]{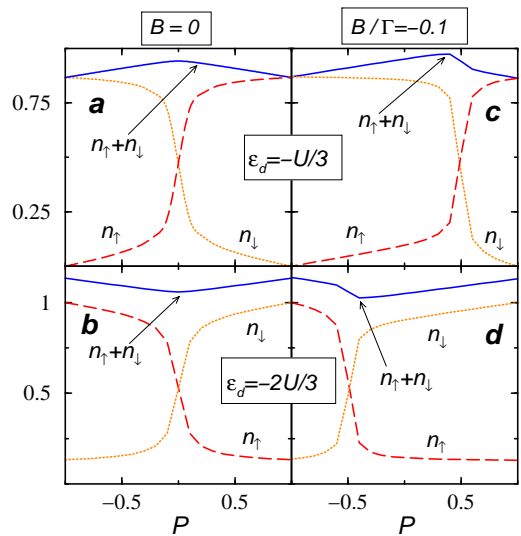}
\caption{(a)-(d): Spin-dependent occupation $n_\s$ of the local
level as a function of the leads' spin polarization $P$ for
$\e_d=-U/3$ [(a) and (c)] and $\e_d=-2U/3$ [(b) and (d)](related
by the particle-hole symmetry) for $B=0$ (left column) and $B=-0.1
\; \Gamma$ (right column). For finite spin-polarization $P\neq 0$
the condition $n_{\up}=n_{\down}$ can only be obtained by an
appropriately tuned magnetic field, as shown in (c) and (d).
Parameters: $U=0.12 \; D_0$, $\Gamma=U/6$.}
 \label{fig:Fig_A_1}
\end{figure}

Consequently we start our numerical analysis by computing the
spin-resolved dot occupation $ n_{\sigma} \equiv \langle
\bar{n}_{\sigma} \rangle $, which is a static property.
Fig.~\ref{fig:Fig_A_1} shows the spin-resolved impurity occupation
as a function of the spin-polarization $P$ of the leads.
Fig.~\ref{fig:Fig_A_1}(a) and (c) corresponds to a gate-voltage of
$\e_{\rm d}=-U/3$ (where the total occupation of the system
$n_{\uparrow}+n_{\downarrow}< 1$) whereas the second line to
$\e_{\rm d}=-2U/3$ (with $n_{\uparrow}+n_{\downarrow}> 1$). The
total occupation of the system $n_{\uparrow}+n_{\downarrow}$
decreases (increases) for $\e_{\rm d}=-U/3$ ($\e_{\rm d}=-2U/3$)
when the spin-polarization of the leads is finite $P\neq 0$.
Note that both situations, $\e_{\rm d}=-U/3$ and $\e_{\rm
d}=-2U/3$), are symmetric in respect of changing particle into
hole states and vice versa, which is possible only for the leads
with particle-hole symmetry. For the gate voltage $\e_{\rm
d}=-U/2$ there is a particle-hole symmetry in the whole system
leads with a DOS which does share this symmetry even in the
presence of spin asymmetry.
Note, whereas a finite spin polarization leads to a decrease in
$n_{\up}+n_{\down}$ for $\e_{\rm d }>-U/2$, cf.\
Fig.~\ref{fig:Fig_A_1}(a), it results in an increase in
$n_{\up}+n_{\down}$ for $\e_{\rm d }<-U/2$, cf.\
Fig.~\ref{fig:Fig_A_1}(b). Obviously, for $P=0$ and in absence of
an external magnetic field the impurity does not have a preferred
occupation $n_{\up}=n_{\down}$.
Any finite value of $P$ violates this relation:
$n_{\up}>n_{\down}$ for $P>0$ and $\e_{\rm d }>-U/2$ (since
$\delta \epsilon_{\rm d \up}\sim \Gamma_{\down}<\Gamma_{\up}\sim
\delta \epsilon_{\rm d \down}$). For $\e_{\rm d }<-U/2$, on the
other hand, the opposite behavior is found.

\begin{figure}[t!]
\centering
\includegraphics[width=0.85\columnwidth]{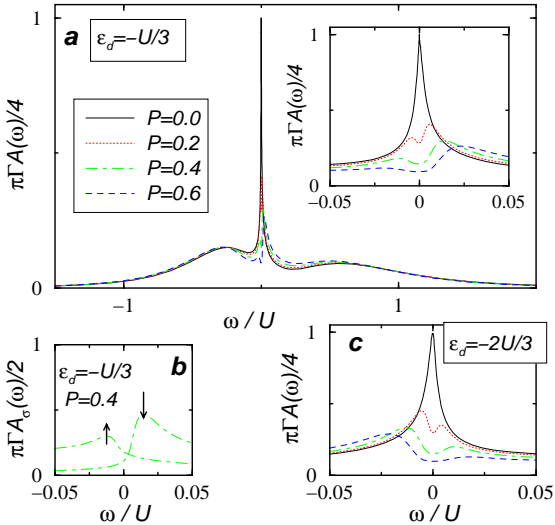}
\caption{Total spectral functions $ A(\omega) = \sum_{\s} A_{
\sigma}(\omega) $ for various values of the spin polarization $P$.
(a) For $\e_{\rm d}=-U/3$ an increase in $P$ results in a
splitting and suppression of the Kondo resonance (see inset). The
effect of finite $P$ on the Hubbard peaks is less significant. The
spin-resolved spectral function $A_{\s}(\o)$, shown in (b),
reveals that $A_{\up}(\o)$ and $A_{\down}(\o)$ differ
significantly from each other for $P\neq 0$. (c) Spectral function
for the same values of $P$ as in (a) but for $\e_d=-2/3 U$ [due to
particle-hole symmetry the results are mirrored as compare to (a),
however with inverted spins]. Parameters: $U=0.12D_0$, $B=0$,
$\Gamma=U/6$.} \label{fig:Fig_A_2}
\end{figure}

The effect on the impurity of a finite leads polarization, namely
to prefer a certain spin species, can be compensated by a locally
applied magnetic field, as shown in Figs.~\ref{fig:Fig_A_1}(c) and
(d).
For $\e=-U/3$ and
$B/\Gamma=-0.1$, cf.\ Fig.~\ref{fig:Fig_A_1}(c), the impurity is
{\it not} occupied by a preferred spin species,
$n_{\up}=n_{\down}$, for $P\sim 0.5$. Due to particle-hole
symmetry the same magnetic field absorbs a lead polarization of
$P\sim -0.5$ for $\e=-2U/3$, cf.\ Fig.~\ref{fig:Fig_A_1}(d). The
magnetic field which restores the condition $n_{\up}=n_{\down}$,
henceforth denote as compensation field $B_{\rm {comp}}(P)$, will
be of particular interest below.

\subsection{Single-particle spectral function}
 \label{sec:spectral}

Using the NRG technique we can access to the spin-resolved
single-particle spectral density $ A_{ \sigma}(\omega,T,B,P)  =
-\frac{1}{\pi} {\rm Im} {\cal G}_{d,\sigma}^R (\omega)$ for
arbitrary temperature $T$, magnetic field $B$, and spin
polarization $P$, where $G_{\sigma}^R (\omega)$ denotes a retarded
Green function.
We can relate the asymmetry in the occupancy, $n_{\up} \neq
n_{\down}$, to the occurrence of charge fluctuations in the dot
and broadening and shifts the position of the energy levels (for
both spin up and down). For $P \neq 0$, the charge fluctuations
and hence level shifts and level occupations become
\emph{spin-dependent}, causing the dot level to split
\cite{martinek1} and the dot magnetization $ n_\uparrow -
n_\downarrow $ to be finite. As a result, the Kondo resonance is
also spin-split and suppressed Fig.~\ref{fig:Fig_A_2}(b),
similarly to the effect of an applied magnetic
field~\cite{Costi_magn_field}. This means that Kondo correlations
are reduced or even completely suppressed in the presence of
ferromagnetic leads.
Note the asymmetry in the spectral function for $P\neq 0$ which
stems from the spin-dependent hybridization $\Gamma_{\s}$. In
Fig.~\ref{fig:Fig_A_2}(b), where the spin-resolved spectral
function is plot, reveals the origin of the asymmetry around the
Fermi energy  of the spectral function. The spectral function
obtained for polarized leads has to be contrasted to that of a dot
asserted to a local magnetic field, where a nearly {\it symmetric}
(perfect symmetry is only for the symmetric Anderson model)
suppression and splitting of the Kondo resonance (around the Fermi
energy) appears.~\cite{Costi_magn_field} In
Fig.~\ref{fig:Fig_A_2}(c) the gate voltage ($\e_{\rm d}=-2U/3$ )
is chosen such that it is particle-hole symmetric to the case
shown in Fig.~\ref{fig:Fig_A_2}(a). Due to the opposite
particle-hole symmetry, the obtained spectral function in
Fig.~\ref{fig:Fig_A_2}(c) is nothing else but the spectral
function shown in Fig.~\ref{fig:Fig_A_2}(a) mirrored around the
Fermi energy.

\begin{figure}[t!]
\centering \includegraphics[width=1.0\columnwidth]{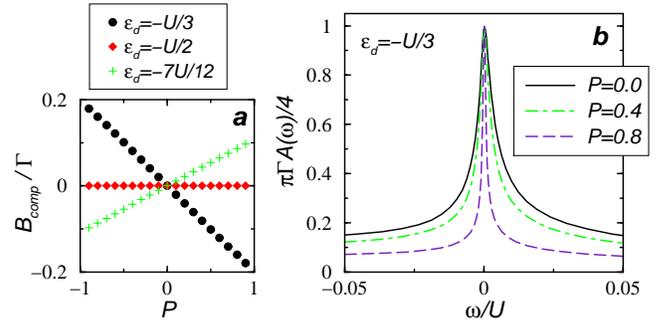}
\caption{(a) Compensation field $B_{\rm comp}(P)$ for different
values of $\e_d$. For flat bands $B_{\rm comp} $ depends linearly
on the leads polarization. At the point where there is
particle-hole symmetry, so for gate voltage ($\e_d=-U/2$), $B_{\rm
comp}(P)=0$ for any value of $P$, i.\ e.\ the spectral function is
not split for any value of $P$ even though $B=0$. (b) Spectral
function for various values of $P$ for $B=B_{\rm comp}$. Note the
sharper resonance in the spectral function, i.\ e.\ a reduced
Kondo temperature $T_{\rm K}$.}
\label{fig:Fig_A_4}
\end{figure}

Fig.~\ref{fig:Fig_A_1}(c) and (d) showed that a finite magnetic
field $B$ can be used to recover the condition $n_{\up}=
n_{\down}$. Indeed, for any lead polarization $P$ a {\it
compensation} field $B_{\rm comp}(P)$ exists at which the impurity
is not preferably occupied by a particular spin species. To a good
approximation $B_{\rm comp}(P)$ has to be chosen such that the
induced spin-splitting of the local level is
compensated~\cite{Bcomp_actual}. One consequently expects a linear
$P$-dependence of $B_{\rm comp}$ from Eq.~(\ref{Bcomp_dep}).
Fig.\ref{fig:Fig_A_4}(a) shows the $P$-dependence of $B_{\rm
comp}$ for various values of $ \epsilon_{\rm d} $ obtained via
NRG. The numerically found behavior can be explained through
Eq.~(\ref{Bcomp_dep}). It confirms that the slope of $B_{\rm
comp}$ is negative (positive) for $\e_{\rm d}>-U/2$ ($\e_{\rm
d}<-U/2$) and that $B_{\rm comp}(P)=0$ for $\e_{\rm d}=-U/2$.

The fact that this occurs simultaneously with the disappearance of
the Kondo resonance splitting suggests that the local spin is
fully screened at $B_{\rm comp}$.
The expectation of an unsplit Kondo resonance in presence of the
magnetic field $B_{\rm comp}$ is nicely confirmed by our numerics
in Fig.~\ref{fig:Fig_A_4}(b) for $\e_{\rm d}=-U/3$. Moreover, the
sharpening of the Kondo resonance in the spectral function upon an
increase in $P$ (indicating a decrease in the Kondo temperature
$T_{\rm K}$) can be extracted from this plot. The associated
binding energy of the singlet (the Kondo temperature $T_{\rm K}$)
is consequently reduced.

\begin{figure}[t!]
\centering \includegraphics[width=1.0\columnwidth]{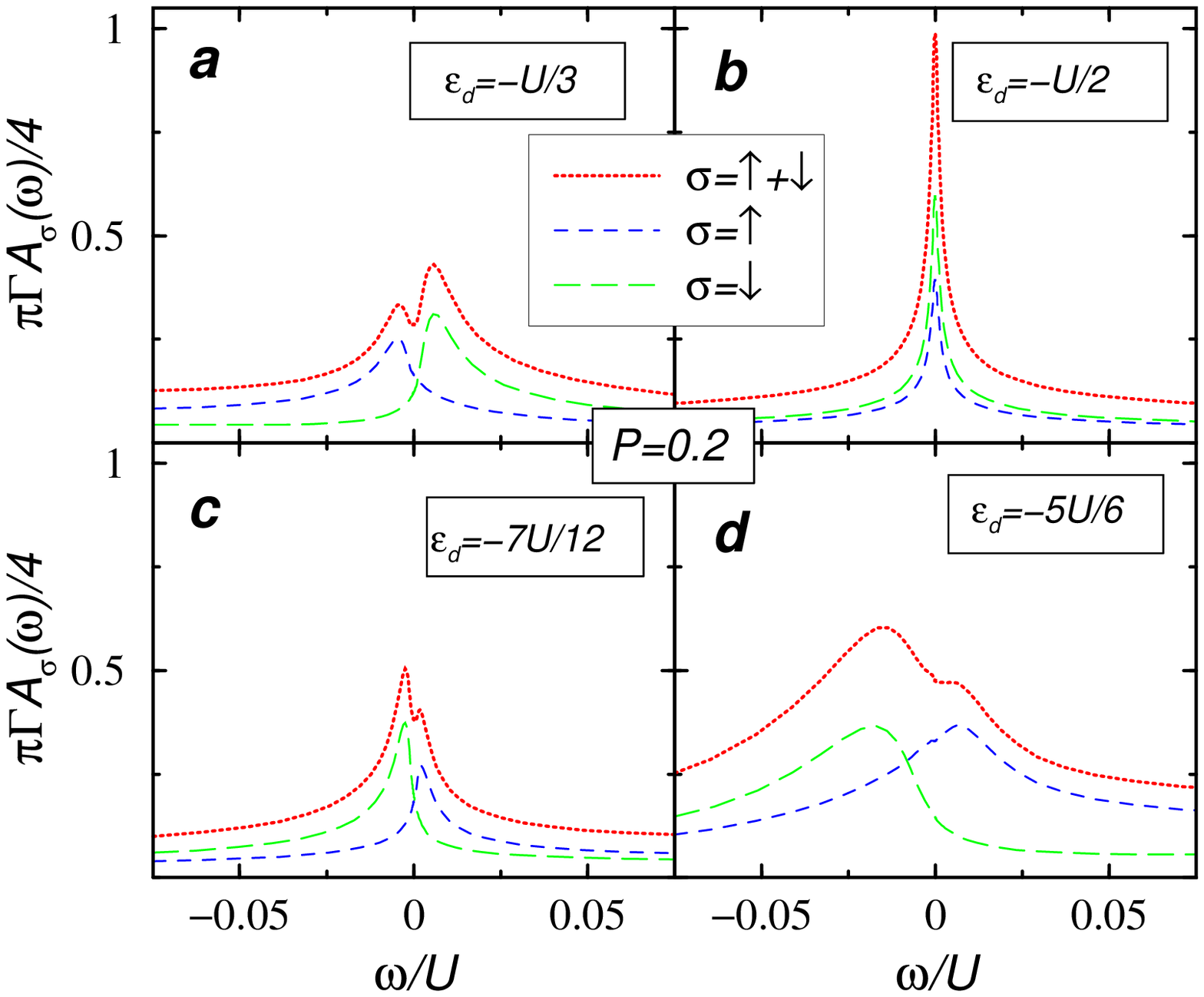}
\caption{Spin-dependent spectral function $A_{\s}(\o)$ for various
values of $\e_{\rm d}$, fixed $P=0.2$, and $B=0$ [(blue) dashed:
$A_{\up}(\o)$, (green) long-dashed: $A_{\up}(\o)$ and (red) dotted
$A(\o)=\sum_{\s}A_{\s}(\o)$]. The splitting between $A_{\up}$ and
$A_{\down}$ changes its sign at $\e_{\rm d}=-U/2$. The spectral
function $A(\o)$ is plot for several values of $P$ for $\e_{\rm
d}=-U/3$ in Fig.~\ref{fig:Fig_A_2} (a). The splitting of the
spectral function $A(\o)$ depends as well on $\e_{\rm d}$ as on
$P$. Parameters: $U=0.12D_0$, $\Gamma=U/6$.}
 \label{fig:Fig_A_3}
\end{figure}
Having demonstrated that the Kondo resonance can be fully recovered
for $B_{\rm comp}$ even though $P\neq 0$ we show that there is also the possibility to recover
the unsplit Kondo resonance via an appropriately tuned gate-voltage.
In Fig.~\ref{fig:Fig_A_3} we plot the spin-resolved spectral function for various
values of $\e_{\rm d}$ for $P=0.2$ and $B=0$.
Note that one can easily identify whether the Kondo resonance is fully
recovered from the positions of $A_{\s}(\o)$ relative to each other.
Since a dip in the total spectral function $A(\o)=\sum_{\s}A_{\s}(\o)$
is not present for a modest shift of $A_{\s}(\o)$ w.\ r.\ t.\ each other,
we identify an unsplit Kondo resonance with perfectly aligned
spin-resolved spectral functions henceforth.

Clearly one can identify from Fig.~\ref{fig:Fig_A_3}
that the spectral function is split for any value $\e_{\rm d}\neq -U/2$.
{\laszlo This splitting} changes its sign at $\e_{\rm d}= -U/2$.
For the particular case $\e_{\rm d}= -U/2$, cf.\
Fig.~\ref{fig:Fig_A_3}(b), the spectral function reaches the
unitary limit. For this gate voltage the spin-resolved spectral
functions $A_{\s}(\o)$ have a different height as predicted from
the Friedel sum rule Eq.~(\ref{eq:A}).

\subsection{Friedel sum rule}
 \label{sec:Friedel}

A direct consequence of the spin-splitting of the local level and
the accompanied $n_{\up}\neq n_{\down}$ for $P\neq 0$ and $B=0$
can be explained by means of the Friedel sum
rule,~\cite{Langreth:1966} an exact $T=0$ relation valid also for
arbitrary values of $P$ and $B$.
This formula relates the height of the spectral function (at the
Fermi energy) and the phase shift of the electrons 
scattered by
the
impurity. As already pointed out spin-polarized
electrodes induce a splitting the local impurity level resulting
in a spin-dependent occupation of the impurity. Thus the knowledge
of the local occupation can be used to anticipate (via the Friedel
sum rule) that the local spectral function becomes spin-split and
suppressed (similar to the presence of a local magnetic field) in
the presence of spin-polarized leads.

%
%
%
%
%
%
%
According to Friedel sum rule,
the height of the spectral function at the Fermi energy $A_\sigma
(0)$ are fully determined by the level occupation $ n_\sigma$
\begin{eqnarray}
{ \phi_\sigma }   =   \pi n_\sigma  , \qquad
 A_{\sigma}(0 ) &
= & { {\rm sin}^2(\pi n_\sigma ) \over \pi \Gamma_{\sigma} } \; ,
 \label{eq:A}
\end{eqnarray}
where $\phi_\sigma$ being the spin dependent phase shift.
This implies a split and suppressed Kondo resonance for $P\neq 0 $
in absence of a magnetic field.

\begin{figure}[t!]
\centering \includegraphics[width=1.0\columnwidth]{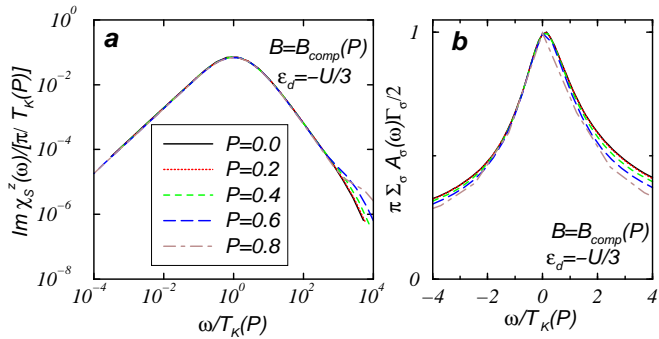}
\caption{The isotropic Kondo effect, accompanied by universal
scaling, can be recovered even for $P\neq 0$ and $B=B_{\rm comp}$
since both the spin-susceptibility $\Im\chi_S^z$ (a) and the
spectral function (b) collapse onto a universal curve. This
statement holds for any value of $\e_d$ with some corresponding
compensation fields as given in Fig.~\ref{fig:Fig_A_4}(a).
Parameters: $U=0.12D_0$, $\e_d=-U/3$, $\Gamma=U/6$.}
 \label{fig:Fig_A_5}
\end{figure}
%
An equal spin occupation, $
n_\uparrow =  n_\downarrow $, 
%
%
can be obtained only for an appropriate external magnetic
field $ B = B_{\rm comp}$. For the latter, in the local moment
regime ($n\approx 1$) we have $n_\uparrow = n_\downarrow \approx
0.5$, so that  $\phi_\uparrow = \phi_\downarrow \approx \pi/2$,
which implies that the peaks of $A_\uparrow$ and $A_\downarrow$
are aligned. Thus, the Friedel sum rule clarifies why the magnetic
field $B_{\rm comp}$ at which the splitting of the Kondo resonance
disappears, coincides with that for which $n_\uparrow =
n_\downarrow$.
Eq.~(\ref{eq:A}) indicates a new feature namely that the amplitude of the 
Kondo resonance becomes spin dependent (see also the discussion in
Sec.~\ref{sec:Conductance}).

It is important to point out that Eq.~(\ref{eq:A}) is valid only
for the leads with particle-hole symmetry (see
Ref.\onlinecite{Hewson-book}). For arbitrary DOS shape it is
possible to generalize the Friedel sum rule.

\subsection{Spin-spectral function - Kondo temperature}
 \label{sec:Spin_spectra}

\begin{figure}[t!]
\centering \includegraphics[width=1.0\columnwidth]{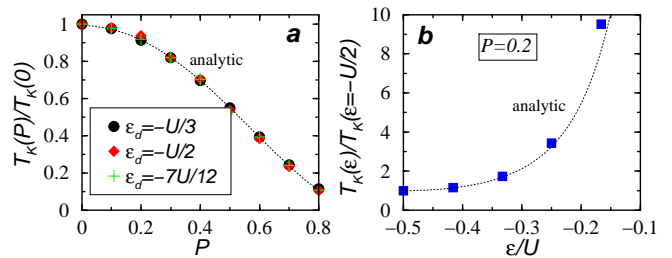}
\caption{(a) Spin polarization dependence of the Kondo temperature
$T_{\rm K}(P)$ [obtained by applying an appropriate $B=B_{\rm
comp}(P,\e_{\rm d})$]. The functional dependence of $T_{\rm K}(P)$
found via a poor man's scaling analysis (dotted line), see Eq.(6)
of Ref.~\onlinecite{martinek1}, is confirmed by the NRG analysis.
The plot confirms that a dot tuned to the local moment regime,
$\sum_{\s} n_{\s} \simeq 1$, shows the universal $P$-dependence of
$T_{\rm K}(P)$, for arbitrary values of $\e_{\rm d}$. (b) The
$\e_{\rm d}$-dependence of $T_{\rm K}$ for fixed $P=0.2$ at
$B=B_{\rm comp}$. As long as the impurity remains in the local
moment regime Haldane's formula~\cite{Haldane} for $T_{\rm K}$
(dotted line) properly describes $T_{\rm K}(\e_{\rm d})$ even
though the leads have a finite spin polarization. Parameters:
$U=0.12 \; D_0$, $\Gamma=U/6$.} \label{fig:Fig_A_6}
\end{figure}

We already stated above that the Kondo temperature $T_{\rm K}$
decreases when the leads polarization $P$ is increased. Obviously,
$T_{\rm K}$ has to vanish when we are dealing with fully
spin-polarized leads $|P|=1$.
To obtain $T_{\rm K}(P)$ we calculated the imaginary part of the
quantum dot spin spectral function
\be
 \chi_S^z(\omega)=\mathcal{F}\{i\Theta(t)\langle
[S_z(t),S_z(0)]\rangle\}
 \label{eq:spincorrelation}
 \ee
 ($\mathcal{F}$ denotes the Fourier
transform), see also Fig.~\ref{fig:Fig_A_5}(a), in presence of the
appropriate $B_{\rm comp}(P,\e_{\rm d})$ and identified the
maximum in $\Im \chi_S^z$ with $T_{\rm K}(P)$.
Fig.~\ref{fig:Fig_A_6}(a) shows $T_{\rm K}(P)$ normalized to
$T_{\rm K}(P=0)$ for different values of $\e_{\rm d}$. As shown in
Ref.[\onlinecite{martinek1,martinek2}] the functional dependence
of $T_{\rm K}(P)$ can be nicely explained within the framework of
poor man's scaling. Note that the decrease in $T_{\rm K}$ is
rather weak for a modest value of $P$. In
Fig.~\ref{fig:Fig_A_5}(b) we plot, additionally to $T_{\rm K}(P)$,
the gate-voltage dependence of $T_{\rm K}(\e_{\rm d})$ normalized
to $T_{\rm K}(\e_{\rm d}=-U/2)$ (where the Kondo temperature is
minimal) for fixed $P=0.2$. In presence of $B_{\rm comp}(P,\e_{\rm
d})$ this functional dependence of the Kondo temperature can
analytically be described via Haldane's formula~\cite{Haldane} for
the Kondo temperature of a single level dot coupled to unpolarized
leads, $T_{\rm K}=\half\sqrt{U\Gamma}e^{\pi\e_d(\e_d+U)/\Gamma
U}$. This fact is another manifestation that indeed the usual
Kondo effect can be recovered in presence of spin-polarized leads,
given the appropriate $B_{\rm comp}(P,\e_{\rm d})$ is applied.

Having shown that the coexistence of spin-polarization in the
leads and the Kondo effect is possible (given a magnetic field $B$
is tuned appropriately at a given  gate-voltage $\e_{\rm d}$), we
plot the properly rescaled spin-susceptibility $\Im \chi_S^z$, see
Fig.~\ref{fig:Fig_A_5}(a), and spectral function, see
Fig.~\ref{fig:Fig_A_5}(b), confirming that the isotropic Kondo
effect is recovered. The expected collapse of $\Im \chi_S^z$ and
$\sum_{\s}\Gamma_{\s} A_{\s}(\o)$ onto a universal curve for
temperatures below $T_{\rm K}$ [note that $T_{\rm K}$ depends on
$P$, see e.\ g.\ Fig.~\ref{fig:Fig_A_4}(b)] is confirmed by our
numerical results. For energies much bigger that $T_{\rm K}$,
$\omega\gg T_{\rm K}$, universal scaling is lost and the curves
start to deviate from each other.

\subsection{Conductance}
 \label{sec:Conductance}

\begin{figure}[t!]
\centering \includegraphics[width=1.0\columnwidth]{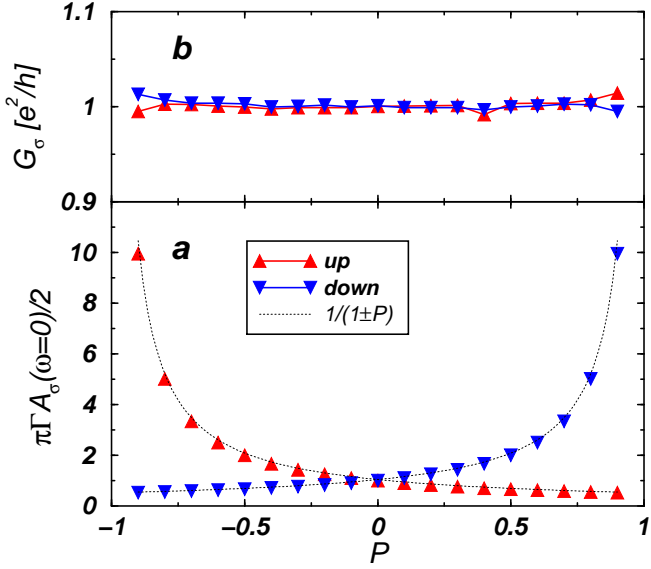}
\caption{(a) Height of the spin-resolved spectral function
$A_{\s}(\omega=0)$ as a function of $P$. The dashed line shows the
$1/(1\pm P)$ dependence as expected from the Friedel sum rule,
Eq.~(\ref{eq:A}). (b) Spin-resolved conductance $G_{\s}$ as
obtained from Eq.~(\ref{spin_res_Cond}). This plot confirms the
expectation based on the Friedel sum rule that the spin-resolved
conductance should be independent of the spin species,
consequently it serves as a consistency check of our numerics.}
\label{fig:4_PRL}
\end{figure}

The knowledge of the spectral function enables us to compute
quantities which are experimentally accessible, such as the linear
conductance. To investigate the question whether spin-polarized
ferromagnetic leads introduce a spin-dependent current we compute
the spin-resolved conductance $G_{\sigma}$,
 \be
 \label{spin_res_Cond}
 G_{\sigma} = {e^2 \over \hbar } \frac{2
\Gamma_{\rm L \sigma } \Gamma_{\rm R  \sigma } }{(\Gamma_{\rm L
\sigma } + \Gamma_{\rm R \sigma } ) }
 \int_{- \infty } ^{ \infty}
d \omega
A_{\sigma}(\omega) \left(-{ \partial f( \omega ) \over
\partial \omega } \right)
\ee
with $f( \omega )$ denoting the Fermi function (remember that we
choose $ \Gamma_{\rm L \sigma } = \Gamma_{\rm R \sigma } $). The
total conductance $G$ is nothing else but the sum of the
spin-resolved conductances $G=\sum_{\s}G_{\sigma}$. Based on the
Friedel sum rule, Eq.~(\ref{eq:A}), we expect the height of the
spin-resolved spectral function at the Fermi energy $A_{\s}(0)\sim
1/\Gamma_{\s} \sim 1/(1\pm P)$. This expectation is nicely
confirmed by our numerical results in Fig.~\ref{fig:4_PRL}(a). As
the $T=0$ conductance is given by the product of $A_{\s}(0)$ and
$\Gamma_{\s}$ the spin-resolved conductance $G_{\sigma} \sim
\Gamma_{\s} A_{\s}(0) $ becomes $P$ independent. The results of
the NRG calculation are shown in Fig.~\ref{fig:4_PRL}(b)
confirming this expectation. We conclude that, even though one is
dealing with spin-polarized leads, it is {\it not } possible to
create a spin current via spin polarized leads the current remains
spin independent. For the {\emph antiparallel} alignment due to
the DOS mismatch, the conductance will be reduced below $G_0=2
e^2/h$ limit.


\section{Flat bands with Stoner splitting}
 \label{sec_flatband_shift}

In ferromagnetic leads, additional to a finite lead spin
polarization $P$ at the Fermi surface, a splitting between the
spin-$\up$ and spin-$\down$ bands $\Delta_{\sigma}$ (Stoner
splitting), which is effect of the effective exchange magnetic
field in the leads, might appear. Taking this effect into account
we do {\it not} expect the splitting of the Kondo resonance to
disappear at $\e_{\rm d}=-U/2$ as in Fig.~\ref{fig:Fig_A_3}(b),
rather at a different value of $\e_{\rm d}$ determined by the
shifts $\Delta_{\sigma}$.
This is due to the fact that the Stoner splitting breaks the
particle-hole symmetry in the electrodes.
Consequently we shall consider flat, spin-polarized leads (of
polarization $P$), which are shifted relative to each other by an
amount  $\Delta_{\sigma}$ in this section. Since particle-hole
symmetry is lost for this band structure in the leads we expect
neither the splitting of the Kondo resonance, see e.\ g.\
Fig.~\ref{fig:Fig_A_3}, nor the compensation field $B_{\rm comp }$
to be symmetric around $\e_{\rm d}=-U/2$, in contrast to
Section~\ref{sec_flatband_noshift}.
\begin{figure}[t!]
\centering
\includegraphics[width=0.350\columnwidth]{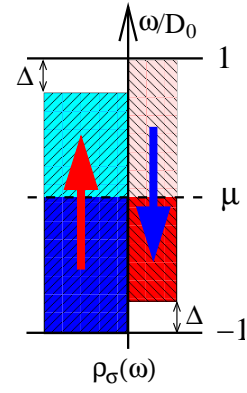}
\caption{The DOS in case of flat bands with the Stoner splitting
$\Delta$ and a finite spin polarization $P$. In general the shifts
of the $\up$ and $\down$ bands are unrelated to each other,
$\Delta_{\down}\neq -\Delta_{\up}$. In this section, however, we
assume $\Delta_{\down}= -\Delta_{\up}$.}
 \label{fig:DOS_Flat_constshift}
\end{figure}

To quantify the statement that a shift $\Delta_{\sigma}$
introduces an finite exchange field and splitting, in this section
we analyze the particular case $\Delta_{\down}=
-\Delta_{\up}=\Delta$, sketched in
Fig.~\ref{fig:DOS_Flat_constshift}. Note that the leads DOS
structure discussed in this section~\cite{use_band} already
requires the use of the extended NRG scheme, explained in
Section~\ref{sec_NRG_mapping}. In particular, spin-dependent
on-site energies along the Wilson chain $\e_{n\s}$, as explained in
Section~\ref{sec_NRG_mapping} and Appendix~\ref{aver_hybr} and
\ref{app_CB}, need to be determined to solve for the leads band
structure shown in Fig.~\ref{fig:DOS_Flat_constshift}.

The studies summarized in Section~\ref{sec_flatband_noshift}
revealed that the spin-resolved impurity occupation $n_{\sigma}$
is the key quantity in the context of a dot contacted to
spin-polarized leads.
\begin{figure}[t!]
\centering \includegraphics[width=1.0\columnwidth]{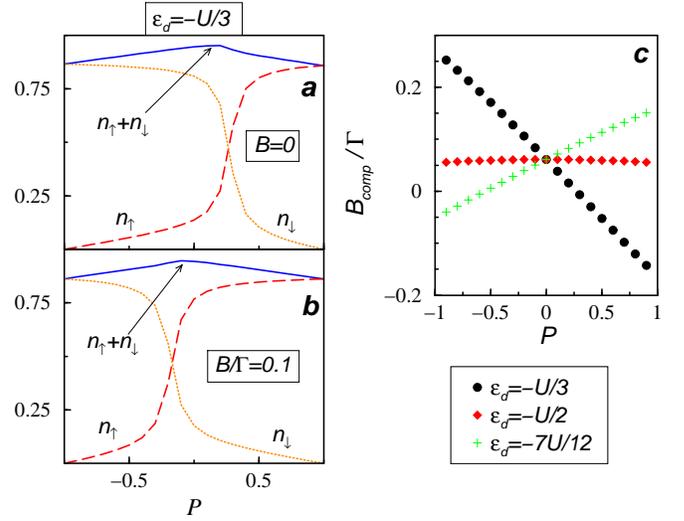}
\caption{Spin polarization $P$ dependence of the local occupation
$n_{\s}$ for $B=0$ (a) and  $B=0.1 \; \Gamma$ (b) with $\e_{\rm
d}=-U/3$. The Stoner-splitting $\Delta$ even in the absence of
spin polarization of the leads, $P=0$, introduces an effective
exchange field which splits the local level $n_{\down}>n_{\up}$ in
absence of an external magnetic field $B=0$. The spin polarization
dependence of $B_{\rm comp}$ for various values of $\e_d$ is shown
in (c). To a good approximation the Stoner-splitting introduces a
constant exchange field (here: $B_{\rm comp}/\Gamma\approx
0.061$), therefore it does nothing else but shifting the $B_{\rm
comp}$-dependence of section~\ref{sec_flatband_noshift}.
Parameters: $U=0.12\l(\frac{D}{D_0}\right)D_0$, $\Gamma=U/6$,
$\Delta_{\down}=-\Delta_{\up}=0.10\l(\frac{D}{D_0}\right)D_0$.}
 \label{fig:Fig_B_1}
\end{figure}
Following this finding we plot $n_{\s}$ vs.\ $P$ for leads with
$\Delta=0.1\l(\frac{D}{D_0}\right)D_0$ and $\e_{\rm d}=-U/3$ in
Fig.~\ref{fig:Fig_B_1}. In contrast to the previous section the
impurity is preferably occupied by $\down$-electrons in the
absence of a magnetic field, see Fig.~\ref{fig:Fig_B_1}(a) even
for $ P= 0 $. A finite magnetic field, e.\ g.\  $B/\Gamma=0.1$
[Fig.~\ref{fig:Fig_B_1}(b)], has the same consequence as described
before, namely the shift the local impurity levels $\e_{\rm d
\s}$.
The $P$-dependence of $B_{\rm comp}$ for the particular band
structure sketched in Fig.~\ref{fig:DOS_Flat_constshift} is shown
in Fig.~\ref{fig:Fig_B_1}(c); it is roughly given by the
compensation field for flat bands, Fig.~\ref{fig:Fig_A_4}(a) where
the same gate-voltages were used, shifted by an effective exchange
field  $ \Delta \epsilon_{\rm d}^{(\rm St)}  $ generated by the
band splittings for $ P=0 $. The value of $B_{\rm Stoner}$ can be
obtained via integrating out those band states of energy $\omega$
which lie in the interval $D\leq |\omega|\leq D_0$ (see
Eq.~(\ref{Bcomp_Stoner_symm}). The value of this effective
exchange field, which is logarithmically suppressed (as can be
shown by perturbative scaling), given by
Eq.~(\ref{Bcomp_Stoner_symm}). Inserting the numbers used in this
section we obtain a $\Delta \epsilon_{\rm d}^{(\rm St)} / \Gamma
\approx 0.060$. The numerical calculation reveals $\Delta
\epsilon_{\rm d}^{(\rm St)}/\Gamma \approx 0.061$ [see
Fig.~\ref{fig:Fig_B_1}(c)], i.\ e.\ the numerical result agrees
reasonably well with the result based on scaling analysis (See
Sec.~\ref{sec:scaling}).

\subsection{Finite temperature: Comparison with experimental data}

\begin{figure}[t!]
\centering \includegraphics[width=\columnwidth]{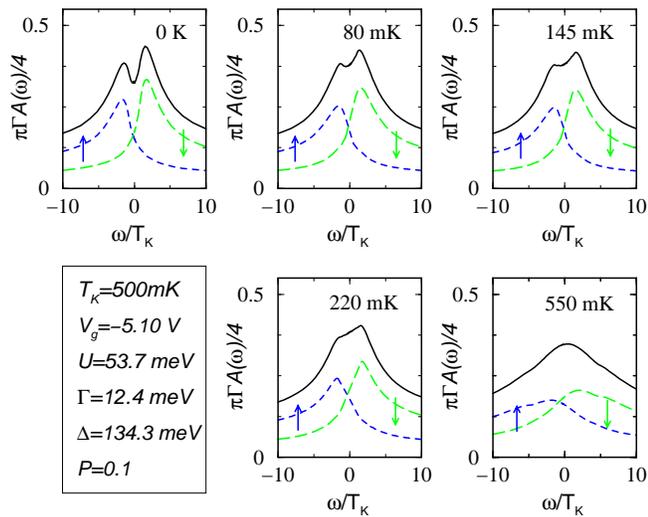}
\caption{Spin-resolved equilibrium spectral function
$A_{\s}(\omega,T,V=0)$ with parameters extracted from
Ref.~\onlinecite{nygaard}. The (blue) dashed lines correspond to
$A_{\up}(\omega,T,V=0)$, the (green) long dashed lines to
$A_{\down}(\omega,T,V=0)$ and the (red) solid ones to the sum of
both, i.\ e.\ $A(\omega,T,V=0)$. (a)-(e) A gradual increase in the
temperature for those temperatures used in
Ref.~\onlinecite{nygaard}. For comparison we plot the $T=0$
spectral function $A(\omega,T=0,V=0)$ in panel (a) as well. Note
that the splitting of the Kondo resonance disappears upon
increasing $T$.}
 \label{Fig_Nygaard_1}
\end{figure}

In a recent experiment of Nyg{\aa}rd~\cite{nygaard} {\it et al.}
an anomalous splitting of the zero bias anomaly in the conductance
induced by the Kondo resonance in {\it absence} of a magnetic
field was observed. In this experiment quantum dot based on single
wall carbon nanotubes (SWNT) contacted to non-magnetic, spin
unpolarized $Cr/Au$ electrodes were used. Those authors related the
observed splitting of the Kondo resonance at zero magnetic field
with the presence of a ferric iron nitrate
nanoparticle~\cite{SWNT_catalyst}, which due electronic tunnel
coupling to the dot introduces a spin-dependent hybridization.

We model the setup of Ref.~\onlinecite{nygaard} by means of a
single-level dot tuned to the local moment regime $\e_{\rm
d\s}=-U/2$ [dashed line in Fig.~1(b) of
Ref.~\onlinecite{nygaard}]. Moreover we used $T_{\rm K}=500 \;
{\rm mK} $~\cite{Marcus_private}, which is in roughly the value of
$T_{\rm K}$ observed in Ref.~\onlinecite{nygaard}, as an input
parameter and extracted from this value $U/\Gamma=4.3$.
Since a splitting of the Kondo resonance was observed for $\e_{\rm
d\s}=-U/2$, see e.g. Fig.~1(b) of Ref.[\onlinecite{nygaard}], a
particle-hole asymmetry must exist to achieve a splitting.
We try to model this effect by using flat band structure and by
introducing the Stoner splitting. We find the best agreement
between the experimentally observed splitting of the Kondo
resonance for $\Delta/U=2.5$. The presence of the nanoparticle
introduces a spin-dependent hybridization which we parameterize
via the leads spin polarization $P$.

The comparison of the $dI/dV$ vs.\ $V$ characteristics of
Ref.[\onlinecite{nygaard}] with the theoretical result turns out
to be rather complicated as it involves the {\it nonequilibrium}
spectral function $A(\omega,T,V)$ which itself is not known how to
compute accurately.~\cite{Rosch,Sindel_frequ}
It is possible to compare qualitatively the splitting in the
non-equilibrium conductance obtained in the experiment to the
single particle spectral function. Due to the splitting in the
particle spectral function for the dot, that is strongly related
with the low-bias conductance measurements we find close
similarity between both of them.
When we approximate the experimental results of $dI/dV$ with
$A(\omega,T,V=0)$ we achieve the best agreement between theory and
experiment for $P=0.1$. For these parameters ($\e_{\rm d\s}=-U/2$,
$U/\Gamma=4.3$, $\Delta/U=2.5$ and $P=0.1$) we compute the
temperature dependent equilibrium spectral function
$A(\omega,T,V=0)$ presented in Fig.~\ref{Fig_Nygaard_1}.
%

\begin{figure}[t!]
\centering \includegraphics[width=\columnwidth]{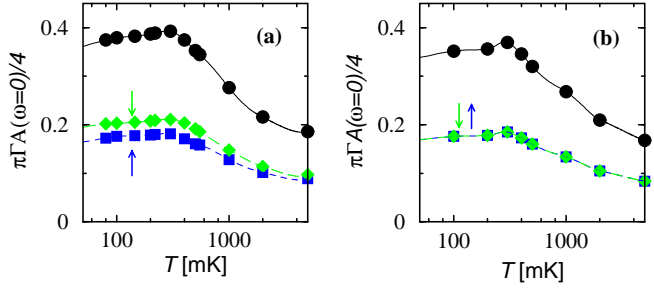}
\caption{Linear conductance as a function of temperature (a) for
the model studied in this section and (b) for the case of a dot
coupled to normal spin unpolarized leads ($P=0$ and $\Delta = 0$)
but in presence of a magnetic field $B=0.02 U$. In both cases a
plateau in the linear conductance around $T_{\rm K}$ can be
observed. Whereas the spin-resolved linear conductance is
degenerate in case (b) (since it is the symmetric Anderson model
with particle-hole symmetry), this is not the case for (a).
Parameters as in Fig.~\ref{Fig_Nygaard_1}, besides $P=\Delta=0$ in
case (b).}
 \label{Fig_Nygaard_3}
\end{figure}

The numerical findings shown in Fig.~\ref{Fig_Nygaard_1}
qualitatively confirm the behavior found in the experiment of
Nyg{\aa}rd and collaborators, namely vanishing splitting in the
$dI/dV$ curves upon with increasing temperature.
%
%
In the experiment, however, the differential conductance $dI/dV$
turns out to be asymmetric around the Fermi energy [see
Fig.~2(b)-(e) in Ref.~\onlinecite{nygaard}]. We attribute the
asymmetry in the $dI/dV$ curve with by the asymmetry in the
coupling between left and right leads.

Finally we want to compare the temperature dependence of the
linear conductance $G$ found in Ref.[\onlinecite{nygaard}] (see
Fig.~2(a) of Ref.[\onlinecite{nygaard}]), with the one based on
the model described above. In absence of a magnetic field $B=0$
and source-drain voltage $V=0$ a plateau in $G$ was found around
$T\approx 500 \; {\rm mK }$ in the experiment. Note that the
calculation of $G$ does not involve the nonequilibrium spectral
function so we can calculate it exactly. In
Fig.~\ref{Fig_Nygaard_3}(a) we show the theoretical curve of the
spin-resolved conductance $G_{\s}$ obtained via
Eq.~(\ref{spin_res_Cond}) for the model discussed in this section.
In agreement with the experiment a plateau in the linear
conductance $G$ is found around $T_{\rm K}$.

\begin{figure}[t!]
\centering \includegraphics[width=0.4\columnwidth]{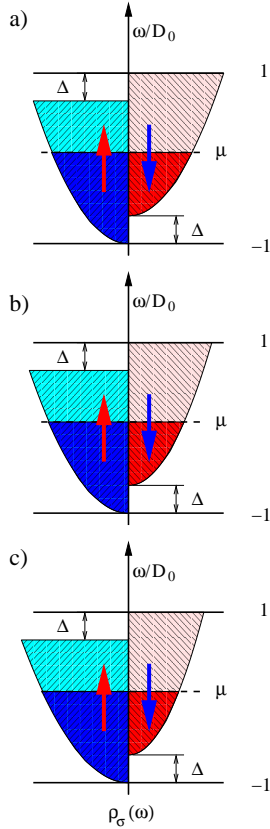}
\caption{ The parabolic density of states (typical for $s$
electron band) given by Eq.~(\ref{DOS}) with the same Stoner
splitting $\Delta = 0.3 D$ but with an additional spin asymmetry
$Q$ defined in the text. Here $Q =0.0 $ (a), $0.1$ (b), and $0.3$
(c).}
 \label{fig:Fig_C_0}
\end{figure}

For comparison we also plot the linear conductance of a dot
contacted to normal spin unpolarized leads (i.\ e.\ flat bands
with $P=0$ which are not shifted relative to each other) for
increasing temperature, however in {\it presence} of a finite
magnetic field $B=0.02 \;U$ in Fig.~\ref{Fig_Nygaard_3}(b). Such a
magnetic field might e.\ g.\ be inserted in the system via the
presence of the ferromagnetic nanoparticles. Also such a scenario
leads to a plateau in $G$. The value of the associated magnetic
field, however, is rather big $B=0.02 \; U$ corresponding to
$B\approx 1$~T [here we assumed $g=2.0$ for the conduction
electrons in the nanotube (see also Ref.~\onlinecite{nygaard})].
According to Ref.~\onlinecite{nygaard} the latter scenario can
therefore {\it not} explain the observed behavior, since one can
argue that such a large magnetic field is hardly to believe can
exist in this experiment (see in particular the footnote 13 of
Ref.[\onlinecite{nygaard}]).

Consequently, the presence of ferromagnetic leads might explain
the experimental observation. We can, however, not exclude other
possible mechanisms - such as the existence of an effective
magnetic field - which might lead to the observed behavior,
however, only form special geometry, where the strong stray
magnetic field is possible.

\section{Band structure with arbitrary shape}
 \label{sec_arb_bands}

\begin{figure}[t!]
\centering \includegraphics[width=0.6\columnwidth]{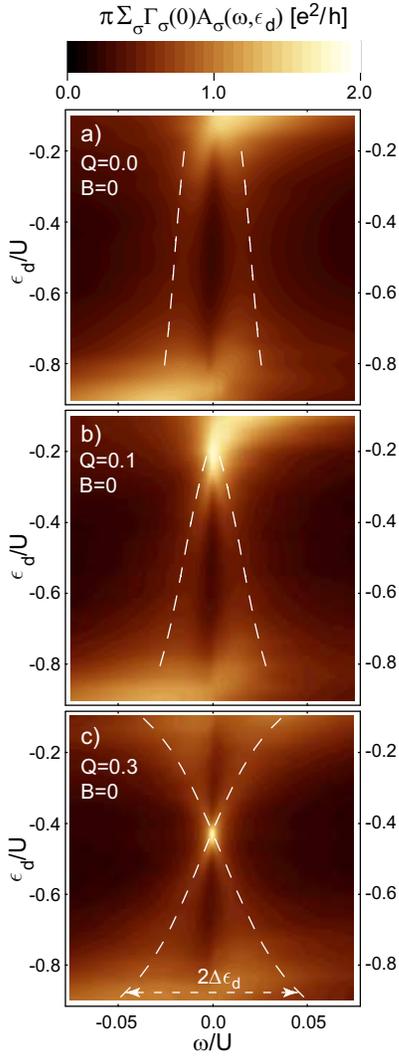}
\caption{Color scale plot of the (rescaled) spectral function
$\sum_{\sigma}\Gamma_{\sigma}A_{\sigma}(\omega)$: the dashed lines
mark the Kondo resonance for the leads band-structure as shown in
 Fig.~\ref{fig:Fig_C_0} in absence of an external magnetic
field. Whereas panel (a)-(c) correspond to a quantum dot in
absence of an external magnetic field, in panel (d)-(f) an
external field is applied.} \label{fig:Fig_C_1}
\end{figure}

\begin{figure}[t!]
\centering \includegraphics[width=0.6\columnwidth]{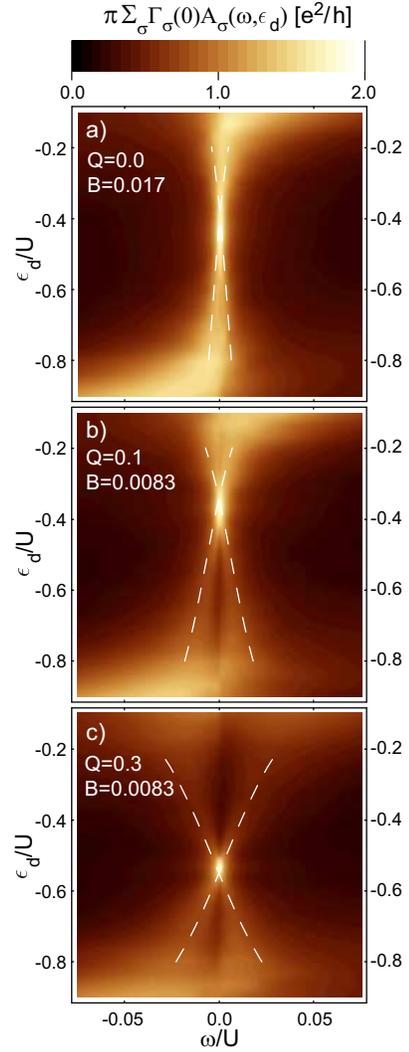}
\caption{Color scale plot of the (rescaled) spectral function
$\sum_{\sigma}\Gamma_{\sigma}A_{\sigma}(\omega)$: the dashed lines
mark the Kondo resonance for the leads band-structure as shown in
Fig.~\ref{fig:Fig_C_0}. In contrast to  Fig.~\ref{fig:Fig_C_1} an
external field is applied. As explained in
Section~\ref{sec_flatband_noshift} a local magnetic field shifts
the spin-resolved local levels relative to each other leading to a
change in the gate voltage where the compensation
($n_{\up}=n_{\down}$) takes place.}
 \label{fig:Fig_C_2}
\end{figure}

\begin{figure}[t!]
\centering \includegraphics[width=0.6\columnwidth]{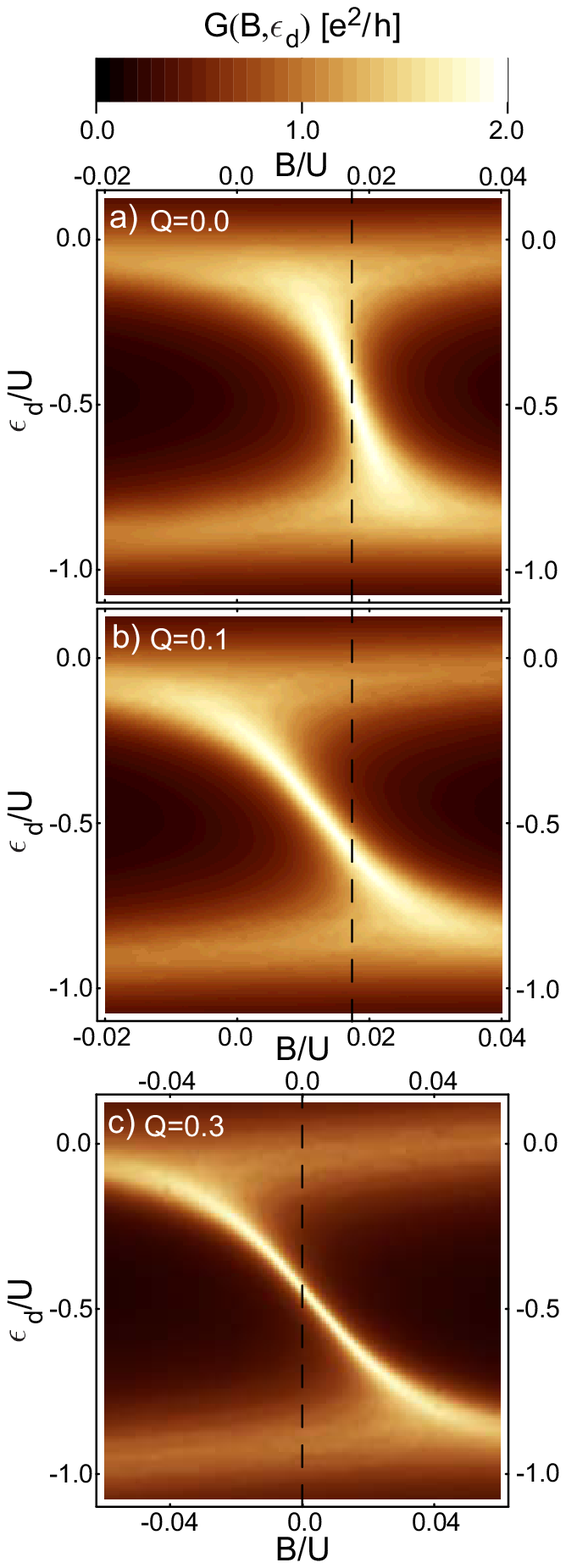}
\caption{ Color scale plot of the linear conductance $G$ as a
function of gate voltage $\e_{\rm d}$ and magnetic field $B$ for
the band-structures corresponding to Fig.~\ref{fig:Fig_C_0}(a)-(c)
[i.\ e.\ (a) corresponds to $Q=0$, (b) to $Q=0.1$ and (c) to
$Q=0.3$]. The black dashed line indicates cross-sections for which
the amplitude is plotted in Fig.~\ref{fig:Fig_C_5}(b).
Note the charging resonances around $\e_{\rm d}\approx 0$ and
$\e_{\rm d}\approx -U$ and the tilted resonance (due to the Kondo
effect) for $-U\lesssim \e_{\rm d} \lesssim 0$.}
 \label{fig:Fig_C_3}
\end{figure}

Flat bands, studied in the previous sections, are only a poor
approximation of a realistic band-structure in the
leads.~\cite{Papaconst}
We complete our study by analyzing the effect of an {\it
arbitrary} leads DOS. Consequently we are considering leads with a
DOS which is energy- and spin-dependent, and additionally contains
a Stoner splitting in this section. As outlined in the previous
section the 'generalized' NRG-formalism introduced in
Section~\ref{sec_NRG_mapping} has to be used in this case as well.
In particular, we study the effect of gate voltage variation on
the spin-splitting of the local level of a quantum dot attached to
ferromagnetic leads. We show how the gate voltage can control the
magnetic properties of the dot. A similar proposal, namely to
control the magnetic interactions via an electric field, was
recently made in gated structures.~\cite{ohno}

In an analysis of several types of band structures we found three
typical gate voltage dependences of the Kondo resonance. To be
more specific, to illustrate this behavior we used a square-root
shape DOS or parabolic band (as for free s-type
electrons)~\cite{free_elec_DOS} with the Stoner splitting
$\Delta$~\cite{yosida}, and some additional spin asymmetry $ Q $
\be
 \label{DOS}
 \rho_{\sigma}(\omega)=
\frac{1}{2}{3 \sqrt{2} \over 8} D^{-3/2} (1+\sigma Q)\sqrt{\omega
+ D + \sigma \Delta/2}
 \ee
depicted in Fig.~\ref{fig:Fig_C_0} [$\sigma \equiv 1(-1) $ for $
\uparrow(\downarrow) $]. This example turns out to encompass
 three typical classes mentioned above. Note that we restricted the
DOS in Eq.~(\ref{DOS}) $\omega \in [-D-\sigma \Delta/2 ,D -\sigma
\Delta/2]$ to ensure that we are dealing with a normalized DOS.

The three
typical gate voltage dependences of the Kondo resonance
we found: (i) a
scenario where the splitting of the local level was roughly
independent of gate voltage [Fig.~\ref{fig:Fig_C_1}(a)], (ii) a
scenario with a strong gate voltage dependence of the splitting,
however without compensation [Fig.~\ref{fig:Fig_C_1}(b)] and (iii)
a case with a strong gate voltage dependence of the splitting
including a particular gate voltage where the splitting vanishes
[Fig.~\ref{fig:Fig_C_1}(c)]. Fig.~\ref{fig:Fig_C_2}(a), (b), and
(c) shows the effect of an external magnetic field for
Fig.~\ref{fig:Fig_C_1}(a), (b), and (c) respectively. As already
outlined in the previous sections an external magnetic field can
be used to compensate the lead induced spin-splitting (which
itself depends on gate voltage) and thereby change that gate
voltage where the full Kondo resonance exists, see
Fig.~\ref{fig:Fig_C_1}(c) and Fig.~\ref{fig:Fig_C_2}(c).
%

\begin{figure}[t!]
\centering \includegraphics[width=0.7\columnwidth]{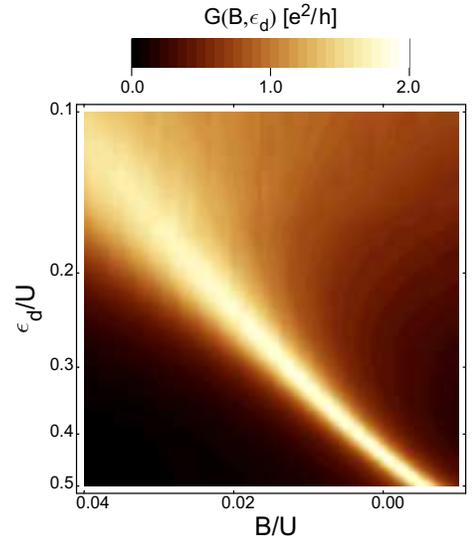}
\caption{ Log-linear version of the Fig.~\ref{fig:Fig_C_3}(c) -
color scale plot of the linear conductance $G$ as a function of
gate voltage $\e_{\rm d}$ and magnetic field $B$ for the
band-structures corresponding to Fig.~\ref{fig:Fig_C_0}(c).  }
 \label{fig:Fig_C_4}
\end{figure}

\begin{figure}[t!]
\centering \includegraphics[width=0.7\columnwidth]{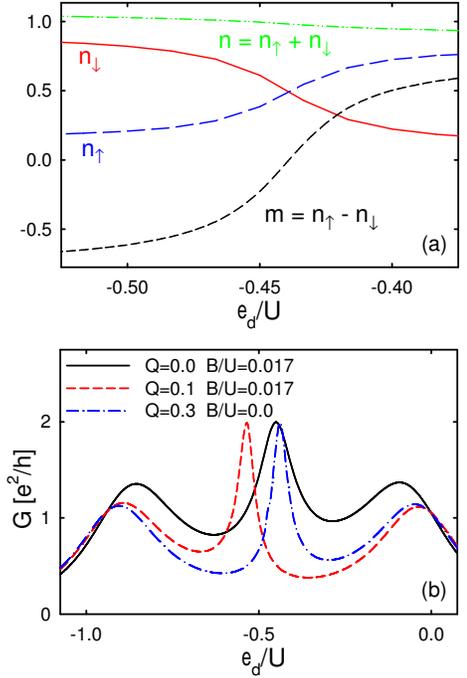}
\caption{ The spin-resolved impurity occupation $ n_{\s}$ for the
DOS reflecting Fig.~\ref{fig:Fig_C_0}(a) and
Fig.~\ref{fig:Fig_C_1}(a) is plot in panel (a). It nicely
illustrates that $\e_{\rm d}$ can be employed to tune the average
impurity magnetization. Since the Kondo resonance is only fully
recovered for $n_{\up}=n_{\down}$ the linear conductance G shows
three peaks in panel (b) for three different situations
corresponding to black dashed lines from Fig.~\ref{fig:Fig_C_3}.}
 \label{fig:Fig_C_5}
\end{figure}

The spitting of the spectral function (see Figs.~\ref{fig:Fig_C_1}
and~\ref{fig:Fig_C_2}) agrees very well with the splitting based
on Haldane's scaling approach~\cite{Haldane} given by the formula
Eq.~(\ref{splitt_gen}). The white dashed line in
Figs.~\ref{fig:Fig_C_1} and~\ref{fig:Fig_C_2} shows prediction of
Eq.~(\ref{splitt_gen}), which fit very well to the numerical
results presented.

In Fig.~\ref{fig:Fig_C_3}(a)-(c) we plot the linear
conductance~\cite{lin_conduc} $G$ for the three band-structures
sketched in Fig.~\ref{fig:Fig_C_0} as a function of gate voltage
$\e_{\rm d}$ and external magnetic field $B$. The two horizontal
maxima in the conductance around $\e_{\rm d}\approx 0$ and
$\e_{\rm d}\approx -U$ correspond to standard Coulomb resonances.
The increased conductance in between, however, is due to the Kondo
effect. Note that, in contrast to the Kondo effect observed in
case of a dot attached to flat and unpolarized leads (where the
Kondo resonance is a vertical line), the Kondo resonance has a
finite slope here. This is due to the gate voltage dependence (and
correspondingly to the compensation field dependence) of the the
spin splitting of the local level [Eq.~(\ref{splitt_gen})].

It is interesting to consider how the Kondo resonance merge with
Coulomb resonances. As one can learn form Eq.~(\ref{Bcomp_dep} )
that the splitting $ \delta \e_{\rm d} $ for the flat band
structure without Stoner splitting shows a logarithmic divergence
for $ \epsilon_0 \rightarrow 0 $ or $ U + \epsilon_0 \rightarrow
0$.
Since any sufficiently smooth DOS can be
 linearized around the Fermi surface, this logarithmic
divergence occurs quite universally, as can be observed in
log-linear versions of Fig.~\ref{fig:Fig_C_3}(c) demonstrated in
Fig.~\ref{fig:Fig_C_4}.
For finite temperature ($ T>0 $) the logarithmic divergence for $
\epsilon_0\rightarrow 0 $ or $\epsilon_0 \rightarrow -U$ is cut
off, $ \Delta \epsilon \simeq - (1 /
 \pi) P \Gamma [ \Psi( 1 / 2 ) + \ln (2 \pi T / U) ] $,
which is also important for temperatures $ T \ll T_{\mathrm K} $.

The finite slope of the Kondo resonance in the linear conductance
plot motivates us to suggest an interesting possibility here: a
dot attached to leads with a DOS as described in Eq.~(\ref{DOS})
and say $Q=0.3$, see Fig.~\ref{fig:Fig_C_3}(c), the gate voltage
can be used to tune the magnetization of the dot. In absence of an
external magnetic field $B=0$ the dot is not occupied by a
preferred spin species for a particular gate voltage (which is
$\e_{\rm d}\approx -U/2$ in this case). Consequently the dot is
preferably occupied with electrons of spin $\s$ for $\e_{\rm
d}\gtrsim -U/2$ and electrons of opposite spin $\bar{\s} $ for
$\e_{\rm d}\lesssim -U/2$. In other words, the average spin on a
dot contacted to leads with such a DOS, $n_{\s}$, and consequently
its magnetization $m=n_{\up}-n_{\down}$ are tunable via an
electric field, which in turn affects $\e_{\rm d}$. This
observation suggests the interesting possibility (similar to
Ref.~\onlinecite{ohno}) to employ an electric field to tune the
magnetic properties of a dot. To compute the gate voltage
dependence of the magnetic field, of course, a detailed knowledge
of the band-structure in the leads is necessary.
Fig.~\ref{fig:Fig_C_5}(a) summarizes this discussion.

Another consequence of the leads band-structure is that the linear
conductance $G$ does not show a plateau when plotted relative to
$\e_{\rm d}$. Instead, we observe a three peak structure in $G$
[see Fig.~\ref{fig:Fig_C_5}(b)] due to the particular gate voltage
at which the full Kondo resonance is recovered.

The theoretical predictions made in this section were done for
particular leads DOS of Eq.~(\ref{DOS}). We used this particular
leads DOS to illustrate the effects on the spin-splitting of the
local level triggered by the band-structure of the leads. To
quantitatively compare our predictions with experiments, however,
a detailed knowledge of the leads DOS is necessary which is in
general very difficult to obtain.

\section{Conclusion} \label{sec_conclusion}

We analyzed the effect of the band structure of the ferromagnetic
leads on the spin splitting of the local level $\e_{\rm d \s}$ of
a quantum dot attached to them. We analyzed the effect of a finite
spin polarization $P$ in case of flat (energy-independent) bands.
We found that the a finite lead polarization results in a spin
splitting, of the local level and the Kondo resonance in a single
particle spectral function, which can be compensated by an
appropriately tuned magnetic field. Given the condition
$n_{\up}=n_{\down}$ is recovered the isotropic Kondo effect is
recovered even though  the dot is contacted to leads with a finite
spin polarization. Additional to this we identified that a finite
Stoner splitting introduces an effective exchange field in the
dot. Finally we explained the consequences of an energy- and
spin-dependent band-structure on the local level. We confirmed
that also in this general case the Kondo resonance can be
recovered. From a methodological point of view we extended the
'standard' NRG procedure to treat leads with an energy- and
spin-dependent DOS.

We thank J.~Barna\'s, T.~Costi, L.~Glazman, W.~Hofstetter,
B.~Jones, C.~Marcus, J.~Nyg{\aa}rd, A.~Pasupathy, D.~Ralph,
A.~Rosch, Y.~Utsumi, and M.~Vojta for discussions. This work was
supported by the DFG under the CFN, 'Spintronics' RT Network of
the EC RTN2-2001-00440, Projects OTKA D048665, T048782, SFB 631,
and by the Polish grand for science in years 2006-2008 as a
research project. Additional support from CeNS is gratefully
acknowledged. L.B. is a grantee of the J\'anos Bolyai Scholarship.
This research was also supported in part by the National Science
Foundation under Grant No. PHY99-07949.

\appendix

\section{Transformation of tunneling Hamiltonian - $\H_{\ell d}$}
 \label{aver_hybr}

The continuous representation of the tunneling Hamiltonian
$\H_{\ell d}$, Eq.~(\ref{eq:Htunneling}),  can be rewritten in
terms of discrete conduction band operators for each spin
component separately by replacing the continuous conduction band
operators $\a_{\o\s}$ by discrete ones, as suggested in
Eq.~(\ref{Fourier_CB})
\begin{eqnarray}
 \H_{\ell d}= \sum_{np\s}\l[d^{\dag}_{\s}
\l(a_{np\s}\int_{\L^{-\l(n+1\right)}}^{\L^{-n}}d\o\; h_{\s}(\o)
\Psi_{np}^{+}(\o)\;+\; \right.\right. \nonumber \\
  \;\;\;\l.\l.
+\; b_{np\s}\int^{-\L^{-\l(n+1\right)}}_{-\L^{-n}}d\o\; h_{\s}(\o)
\Psi_{np}^{-}(\o) \right) + \textrm{h.c.}\right] \: .
 \label{App_tunelling}
\end{eqnarray}
As outlined in Section~\ref{sec_NRG_mapping} we replace the energy
dependent generalized hybridization of the $n$-th logarithmic
interval $h_{\s}(\o)$, $\Lambda^{-(n+1)}\pm\o<\Lambda^{-n}$, by a
constant $h^{\pm}_{n\s}$, defined as
\begin{widetext}
\bea*
 h^+_{n\s}
 &\equiv& \cases{ \frac{1}{d_n}
\int_{\L^{-(n+1)}}^{\L^{-n}}d\o \sqrt{\rho_{\s}(\o)
\l[V_{\s}(\o)\right]^2} \;\;\;&\textrm{if} $\Lambda^{-(n+1)}\leq
\o <\Lambda^{-n}$  \cr 0 &\textrm{else} \cr }
\\
h^-_{n\s} &\equiv& \cases{
\frac{1}{d_n}\int_{-\L^{-n}}^{-\L^{-(n+1)}}d\o \sqrt{\rho_{\s}(\o)
\l[V_{\s}(\o)\right]^2} &\textrm{if} $-\Lambda^{-n}< \o \leq
-\Lambda^{-(n+1)}$  \cr 0 &\textrm{else} \cr }. \eea*
\end{widetext}
Consequently possible integrals in Eq.~(\ref{App_tunelling}), such
as
$\int_{\L^{-\l(n+1\right)}}^{\L^{-n}}d\o\; h_{\s}(\o)
\Psi_{np}^{+}(\o) \propto
\int_{\L^{-\l(n+1\right)}}^{\L^{-n}}d\o\Psi_{np}^{+}(\o) $, give
only for $p=0$ (due to the Riemann-Lebesgue Lemma) a finite
contribution. For the particular choice of constant hybridization
the impurity couples to $s$-waves ($p=0$ modes) only; consequently
we skip the harmonic index $p$ below. With above definitions
Eq.~(\ref{App_tunelling}) simplifies to the following compact form
\be \label{App_tunn_comp} \H_{\ell d}=
\frac{1}{\sqrt{\pi}}\sum_{\s}\l[d^{\dag}_{\s} \sum_{n}
\l(a_{n\s}\gamma_{n\s}^+ +
b_{n\s}\gamma_{n\s}^-\right)\;+\textrm{h.c.}\right], \ee
where
 \begin{eqnarray}
 \gamma_{n\s}^+ & \equiv &
\int_{\L^{-\l(n+1\right)}}^{\L^{-n}}d\o\;\sqrt{\pi\rho_{\s}(\o)
\l[V_{\s}(\o)\right]^2} \nonumber
 \\
\gamma_{n\s}^{-} & \equiv &
\int^{-\L^{-\l(n+1\right)}}_{-\L^{-n}}d\o\; \sqrt{\pi\rho_{\s}(\o)
\l[V_{\s}(\o)\right]^2} \; .
 \label{eq:gammas}
 \end{eqnarray}

Remember that Eq.~(\ref{equal_action}) forces the generalized
dispersion $g_{\s}(\e)$ to be adjusted accordingly for this
particular choice of $h_{\s}(\o)$. It is explained in details in
Appendix~\ref{app_CB}.

\section{Mapping of conduction band onto semi-infinite chain}
\label{app_CB}

Here we outline the important steps to bring the leads Hamiltonian
$\H_{\ell}$, Eq.~(\ref{eq:Hcontinuous}), into the tridiagonal from
introduced in Eq.~(\ref{lead_Wilson_chain}). The replacement of
the continuous conduction band operators $\a_{\o\s}$ by discrete
operators, cf.\ Eq.~(\ref{Fourier_CB}), simplifies $\H_{\ell}$
significantly.~\cite{Sindel_Diss} As shown in
Ref.~\onlinecite{Bulla:1997} the particular choice of the
generalized hybridization function $h_{\s}(\o)\ra h_{n\s}^{\pm}$
(given in Appendix~\ref{aver_hybr}) results in
\be
 \label{H_CB1}
 \H_{\ell}=\sum_{n\s}
\l[\zeta_{n\s}^+a^{\dag}_{n\s}a_{n\s}+\zeta_{n\s}^-b^{\dag}_{n\s}b_{n\s}\right],
\ee where
\begin{eqnarray}
\zeta^+_{n\s}&\equiv& \frac{ \int_{\L^{-(n+1)}}^{\L^{-n}}d\e\;\e\;
\rho_{\s}(\e)} { \int_{\L^{-(n+1)}}^{\L^{-n}}d\e\; \rho_{\s}(\e)}
 \; ,
\\
 \zeta^-_{n\s}&\equiv&
\frac{\int_{-\L^{-n}}^{-\L^{-(n+1)}}d\e\;\e\; \rho_{\s}(\e)}
{\int_{-\L^{-n}}^{-\L^{-(n+1)}}d\e\; \rho_{\s}(\e)} \; .
\label{app_xsi_n_MINUS}
\end{eqnarray}
To achieve the goal to bring $\H_{\ell}$ into tridiagonal form
the tridiagonalization procedure developed by L\'anczos~\cite{Lanczos:1950}
is used.
In general diagonal and off-diagonal
 matrix elements $\e_{n\s}$ and $t_{n\s}$~\cite{Hofstetter:2000_DISS}
need to be computed in the course of the procedure. These matrix
elements can be obtained by demanding the following relation (see
e.g.\ Ref.~\onlinecite{Bulla:2004})
\begin{eqnarray}
 \label{CB_mapping}
&&\sum_{n\s} \l(\zeta_{n\s}^+\;a^{\dag}_{n\s}a_{n\s}+\zeta_{n\s}^-\;
b^{\dag}_{n\s}b_{n\s}\right)= \\
&&\hspace{.5cm}=\sum_{n\s}\l[\e_{n\s} f_{n\s}^{\dag}f_{n\s}+t_{n\s}
\l(f_{n\s}^{\dag}f_{n+1\s}+f_{n+1\s}^{\dag}f_{n\s} \right)\right].
\nonumber
\end{eqnarray}
The spin-dependent coefficients $u_{nm\s}$ and $v_{nm\s}$ of the
single-particle operator [Eq.~(\ref{eq:coef})] (that acts on the
$n$-th site of the Wilson chain) $f_{n\s}$, given by the Ansatz
Eq.~(\ref{Ansatz_f_NRG}), need to be determined recursively for
each spin component separately. Inverting the Ansatz
Eq.~(\ref{Ansatz_f_NRG}) enables one to express the discrete
conduction band operators as
$a_{n\s}=\sum_{m=0}^{\infty}u_{mn\s}f_{m\s}$ and
$b_{n\s}=\sum_{m=0}^{\infty}v_{mn\s}f_{m\s}$, respectively. When
we insert $a_{n\s}$ and $b_{n\s}$ in the l.h.s.\ of
Eq.~(\ref{CB_mapping}) and compare corresponding $f_{n\s}$
operators on both sides of this equation we obtain
\begin{eqnarray}
\label{app_fn_comparison} &&\sum_{m=0}^{\infty} \l(\zeta_{ms}^{+}
u_{nm\s} a_{m\s}^{\dag} +  \zeta_{ms}^{-} v_{nm\s}
b_{m\s}^{\dag}\right) =\nonumber
 \\ &&\hspace{1.0cm}
 =\e_{n\s}
f_{n\s}^{\dag} + t_{n\s} f_{n+1\s}^{\dag}+t_{n-1\s}
f_{n-1\s}^{\dag} \; .
\end{eqnarray}
In particular, Eq.~(\ref{app_fn_comparison}) for $n=0$ gives the
relation for the operators $f_{0\s}$ yields
\begin{eqnarray}
\label{app_f0_comparison}
&&\sum_{m=0}^{\infty}
\l(\frac{\zeta_{m\s}^{+}\gamma_{m\s}^+}{\sqrt{\xi_{0\s}}} a_{m\s}^{\dag}
+  \frac{\zeta_{m\s}^{-}\gamma_{m\s}^-}{\sqrt{\xi_{0\s}}} b_{m\s}^{\dag}\right)
\nonumber\\
&&\hspace{1.5cm}=\e_{0\s} f_{0\s}^{\dag} + t_{0\s} f_{1\s}^{\dag},
\end{eqnarray}
where the corresponding values of $u_{0m\s}$ and $v_{0m\s}$
[Eq.~(\ref{eq:coef})] have already been inserted. Since the
operators $f_{n\s}$ obey Fermi-statistics,
$\{f_{n\s},f_{n'\s'}^{\dag}\}=\d_{nn'}\d_{\s\s'}$, the
anticommutator of the r.h.s.\ of Eq.~(\ref{app_f0_comparison})
with $f_{0\s}$ yields $\{\e_{0\s} f_{0\s}^{\dag} + t_{0\s}
f_{1\s}^{\dag},f_{0\s} \}=\e_{0\s}$. The corresponding
anticommutator of the l.h.s.\ of Eq.~(\ref{app_f0_comparison})
with $f_{0\s}$ finally results in
\be
 \label{app_e_0_result}
 \e_{0\s}=\frac{1}{\xi_{0\s}}\sum_{m}\l[
\zeta_{m\s}^+\l(\gamma_{m\s}^+\right)^2\;+\;
\zeta_{m\s}^-\l(\gamma_{m\s}^-\right)^2\right].
\ee
The initial hopping matrix element $t_{0\s}$ is readily obtained
from the anticommutator $\{\e_{0\s} f_{0\s}^{\dag} + t_{0\s}
f_{1\s}^{\dag},\e_{0\s} f_{0\s} + t_{0\s} f_{1\s}
\}=(\e_{0\s})^2+(t_{0\s})^2$. As $\e_{0\s}$ is known, $t_{0\s}$
can be easily obtained from Eq.~(\ref{app_f0_comparison})
\begin{eqnarray}
\label{app_t_0_result} \l(t_{0\s}\right)^2= \l\{ \sum_m\l[
\l(\zeta_{m\s}^+\right)^2 \l(\gamma_{m\s}^+\right)^2+
\l(\zeta_{m\s}^-\right)^2 \l(\gamma_{m\s}^-\right)^2 \right] -
\right.
 \nonumber \\
 \l. \sum_m\l[ \zeta_{m\s}^+\l(\gamma_{m\s}^+\right)^2
+ \zeta_{m\s}^-\l(\gamma_{m\s}^-\right)^2\right]
\right\}\textrm{\huge{/}} \xi_{0\s}.
\end{eqnarray}
The knowledge of $\e_{0\s}$, $t_{0\s}$, and $f_{0\s}$ now enables
us to extract the coefficients of $f_{1\s}$ from
Eq.~(\ref{app_f0_comparison})
\begin{eqnarray}
u_{1m\s}&=&\frac{\gamma_{m\s}^+}{\sqrt{\xi_{0\s}}t_{0\s}}
\l(\zeta_{m\s}^+ -\e_{0\s}\right) \; ,
  \nonumber \\
v_{1m\s}&=&\frac{\gamma_{m\s}^-}{\sqrt{\xi_{0\s}}t_{0\s}}
\l(\zeta_{m\s}^- -\e_{0\s}\right) \; .
\end{eqnarray}
A generalization of the argumentation above finally allows one to
obtain the spin-dependent on-site energies $\e_{n\s}$ and hopping
matrix elements $t_{n\s}$ for the $n$-th site of the Wilson chain
\begin{eqnarray}
\e_{n\s}&=&\sum_m\l[(u_{nm\s})^2\zeta_{m\s}^+
+(v_{nm\s})^2\zeta_{m\s}^-\right] \label{app_e_n_ITER},
  \\
\l(t_{n\s}\right)^2&=&\sum_m\l[\l(u_{nm\s}\right)^2\l(\zeta_{m\s}^+\right)^2+\l(v_{nm\s}\right)^2\l(\zeta_{m\s}^-\right)^2
\right]
 \nonumber \\&&\;-\;
\l(t_{n-1\s}\right)^2\;-\;\l(\e_{n\s}\right)^2 \; ,
 \label{app_t_n_ITER}
\end{eqnarray}
where the involved coefficients of the single-particle operator $f_{n+1\s}$,
are given as
 \begin{eqnarray}
 u_{n+1m\s} &=& \frac{1}{t_{n\s}}
\l[\l(\zeta_{m\s}^+-\e_{n\s}\right)u_{nm\s}-
t_{n-1\s}u_{n-1m\s}\right] \; ,
 \nonumber  \\
v_{n+1m\s} &=& \frac{1}{t_{n\s}}\l[\l(\zeta_{m\s}^- -
\e_{n\s}\right)v_{nm\s} - t_{n-1\s}v_{n-1m\s}\right] \; .
 \nonumber  \\
 \end{eqnarray}

Note that it is crucial to determine the coefficients $u_{nm\s}$
and $v_{nm\s}$ even though, only the matrix elements along the
Wilson chain ($\e_{n\s}$ and $t_{n\s}$) are finally required. The
numerical solution of the above mentioned equations is rather
tricky: since the band energies are exponentially decaying in the
course of the iteration one has to use reliable numerical routines
(i.e.\ arbitrary-precision Fortran routines~\cite{Bulla:1997}) to
solve for the coefficients $u_{nm\s}$ and $v_{nm\s}$ and the
matrix elements $t_{n\s}$ or $\e_{n\s}$, respectively.


\begin{thebibliography}{99}
\bibitem{Normalization}
We use normalized density of statets in the leads, $\sum_{r\s}
\int_{-\infty}^{\infty} d\o \rho_{r\s}(\o)=1$, throughout this
report.
\bibitem{interaction}
We expect that a finite $s$-electron interaction, $U_s > 0$, will
lead only to a slight renormalization of $T_{\rm K}$.
\bibitem{remark_1}
 Note that we are dealing with $V_{\s}(\o)=V$, i.e. spin- and energy-independent
 tunneling matrix elements, in this section.
\bibitem{DOS_flat_band}
The spin-dependent DOS in case of flat and spin-polarized bands
(with spin-polarization parameter $P$) is given by
$\rho_{\s}=\half \rho (1\pm P)$, where $\rho=\frac{1}{2D}$ is the
corresponding normalized DOS of the full band.
\bibitem{equivalence_scatt_GF_phase}
For a single level Anderson model the phase of the retarded Green's function is
identical to the scattering phase.
\bibitem{Bcomp_actual}
Note that in case of spin-polarized leads the condition $n_{\up}=n_{\down}$
is not identical to the condition of degenerate levels, as $\Gamma_{\up}\neq \Gamma_{\down}$.
\bibitem{use_band}
Note that the appearance of the Stoner splitting makes the use of
the generalized bandwidth $D_0$ ($\neq D$) necessary.
\bibitem{free_elec_DOS}
The DOS $\rho_{\sigma}(\omega)$ is defined as
$\rho_{\sigma}(\omega)=\sum_{k}\delta(\omega-\e_{k\s})$. Since
free electrons have a dispersion $\e_{k\s}=\frac{\hbar^2k^2}{2m}$
one obtains $\rho_{\sigma}(\omega)\sim \int
\sqrt{\omega'}d\omega'\delta(\omega-\omega') = \sqrt{\omega}$.
\bibitem{lin_conduc}
We obtain the linear conductance $G(\e_{\rm d},B) =
\sum_{\s}G_{\s}(\e_{\rm d},B)$, see Eq.~(\ref{spin_res_Cond}),
 by computing the spin-resolved spectral function $A_{\sigma}(\omega=0,\e_{\rm d},B)$
(for $T=0$) for those values of $\e_{\rm d}$ and  $B$ that are
relevant for Fig.~\ref{fig:Fig_C_2}.
\end{thebibliography}
\end{document}